\documentclass[aps,
prx,
10pt,
twocolumn,
superscriptaddress,
nofootinbib,
longbibliography]{revtex4-2}

\usepackage{amsmath,amssymb,amsfonts,amsthm,mathtools}
\usepackage{dsfont}
\usepackage{bbm}
\usepackage{upgreek}
\usepackage{physics}
\usepackage[scaled=1]{helvet}
\usepackage{makecell}
\usepackage{multirow}
\usepackage{graphicx}
\usepackage{lipsum}
\usepackage[table,svgnames,dvipsnames]{xcolor}
\usepackage{soul}
\usepackage[normalem]{ulem}
\usepackage{enumerate}

\usepackage[colorlinks=true,citecolor=Cerulean,linkcolor=RubineRed,urlcolor=Cerulean]{hyperref}

\newcommand{\id}{\mathbbm{1}}

\newcounter{ls}

\newcommand{\out}[1]{{\color{teal}\ifmmode\text{\sout{\ensuremath{#1}}}\else\sout{#1}\fi}}

\makeatletter
\def\@fnsymbol#1{\ensuremath{\ifcase#1\or \dagger \or \ddagger \or \mathsection \or \mathparagraph \or \| \or ** \or \dagger\dagger \or \ddagger\ddagger \else\@ctrerr\fi}}
\makeatother

\newcounter{am}

\newcounter{os}

\newcounter{ad}

\begin{document}

\title{Randomised measurements of a disorder-induced entanglement transition\\in a neutral atom quantum processor}

\date{\today}

\author{Apollonas S. Matsoukas-Roubeas$^{\ast}$  \href{https://orcid.org/0000-0001-5517-0224}{\includegraphics[scale=0.04]{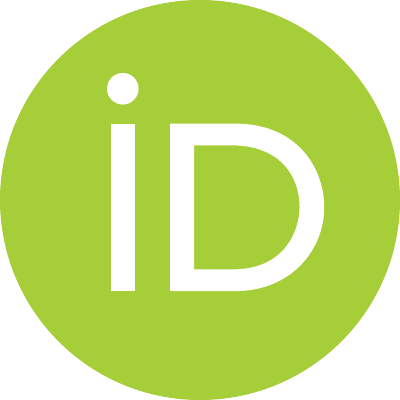}}}
\thanks{Corresponding authors: dag50@cam.ac.uk, sam310@cam.ac.uk}
\affiliation{Cavendish Laboratory, University of Cambridge, Cambridge, CB3 0US UK}
\affiliation{Department of Applied Mathematics and Theoretical Physics,
University of Cambridge, Cambridge CB3 0WA, UK}

\author{Oscar Scholin$^{\ast}$ \href{https://orcid.org/0009-0005-3537-4736}{\includegraphics[scale=0.04]{orcidid.pdf}}}
\affiliation{Cavendish Laboratory, University of Cambridge, Cambridge, CB3 0US UK}
\affiliation{Department of Physics, University of Oxford, Oxford OX1 3PU, UK}

\author{Lucas Sá~\href{https://orcid.org/0000-0002-7359-647X}{\includegraphics[scale=0.04]{orcidid.pdf}}}
\affiliation{Cavendish Laboratory, University of Cambridge, Cambridge, CB3 0US UK}

\author{Arinjoy De~\href{https://orcid.org/0000-0001-9184-8434}{\includegraphics[scale=0.04]{orcidid.pdf}}}
\affiliation{QuEra Computing, Boston, MA, USA}

\author{Majd Hamdan~\href{https://orcid.org/0009-0003-0446-4013}{\includegraphics[scale=0.04]{orcidid.pdf}}}
\affiliation{QuEra Computing, Boston, MA, USA}

\author{Alexei Bylinskii}
\affiliation{QuEra Computing, Boston, MA, USA}

\author{Andrew J. Daley \href{https://orcid.org/0000-0001-9005-7761}{\includegraphics[scale=0.04]{orcidid.pdf}}}
\affiliation{Department of Physics, University of Oxford, Oxford OX1 3PU, UK}

\author{Dorian A. Gangloff \href{https://orcid.org/0000-0002-7100-0847}{\includegraphics[scale=0.04]{orcidid.pdf}}}
\thanks{Corresponding authors: dag50@cam.ac.uk, sam310@cam.ac.uk}
\affiliation{Cavendish Laboratory, University of Cambridge, Cambridge, CB3 0US UK}

\begingroup
\renewcommand\thefootnote{} 
\footnotetext{\hspace{-0.8em}$^\ast$ These authors contributed equally to this work}
\endgroup

\begin{abstract}
The development and spread of entanglement in complex quantum systems is central to exploring many-body phenomena out of equilibrium. Measuring entanglement dynamics can shed light on information scrambling and thermalisation, namely on transitions from many-body quantum chaos to localisation in disordered, interacting systems. In quantum computing systems, entanglement entropy and other nonlinear functions of the density matrix have been recently measured, in particular by using the randomised measurement toolbox. However, it is difficult to implement the required arbitrary unitary rotations on specific subsystems without universal local control. Here we devise and demonstrate the measurement of entanglement entropy in a programmable analogue quantum simulator using a randomised measurement protocol that leverages local energy tuning together with a global field to bypass the need for local gate control.  We implement this on a commercially available neutral-atom quantum simulator, QuEra's Aquila, and use it to show how programmable disorder in the local Hamiltonian parameters leads to a transition from chaotic to localised entanglement dynamics. Given current decoherence times, we clearly resolve disorder-specific, time-dependent entanglement spreading in small systems. Our work extends the utility of programmable analogue quantum simulators, and opens further opportunities for wider randomised measurement toolboxes in a range of other analogue systems. 
\end{abstract}

\maketitle


\section{Introduction}

How quantum information flows between particles in a system reveals essential features, such as whether it will exhibit particular macroscopic phases of matter or yet whether it will thermalise at all. When information flows it leads to nonlocal correlations, making \emph{quantum entanglement} an ideal probe of shared information in a many-particle system~\cite{amico2008}. Directly measuring entanglement in many-body systems is generally challenging, as it is a nonlinear function of the density operator, $\rho$. However, much progress has been made on digital systems with access to local gates, through the randomised measurement toolbox \cite{elben2022}. By applying local unitaries (randomly drawn from a $k$-design) to subsystems of interest and measuring the full system in the computational basis, one can estimate polynomial functions of $\rho$ with exponentially fewer measurements than full tomography --- the randomisation also helps mitigate experimental imperfections in preparing specific subsystem rotations. This toolbox has been applied successfully in a variety of experimental use cases, including estimating the second-order R\'enyi entropy of up to 10-qubit partitions of a 20 trapped-ion system \cite{brydges2019}, comparing cross-platform state fidelities \cite{zhu2022}, and learning the Lindbladian of a 51-qubit system \cite{kraft2025}. Extending this powerful toolbox to a broader range of quantum systems, including analogue quantum simulators, has been an outstanding challenge.

In this work, we engineer a modification to the randomised measurement toolbox, which we tailor to analogue quantum processors without local gate control, to obtain the entropy of entanglement across subsystems. Our protocol is based on a set of unitary rotations that, despite being implemented by a global control field, can approximate universality of local rotations when combined with site-dependent local detuning. We implement our protocol on an array of $^{87}$Rb atomic qubits confined in optical tweezers \cite{wurtz2023}---QuEra's cloud-accessible machine Aquila. This allows us to monitor entanglement spreading as a function of evolution time across a range of programmable Hamiltonians.

In particular, we focus on programming site disorder using qubit-specific energy detunings. We take the transition from quantum chaotic to localised behaviour in disordered, interacting systems both as an example, and as a central concept to understand the parameter regimes in which our analogue quantum simulator can work. More broadly in many-body systems, the rate and extent to which subsystems become entangled over time depend on the structure of the system’s Hamiltonian and initial state. Generic systems with some disorder present, albeit low, are usually found in a so-called quantum chaotic phase, in which entanglement spreads quickly across subsystems and local information is scrambled. The entanglement entropy of subsystems thus increases linearly and reaches a typical value \cite{page1993}. In this situation, the eigenstate thermalisation hypothesis \cite{deutsch1991,srednicki1994,dalessio2016} predicts that the density matrix describing each subsystem approaches a thermal state, even though the global system undergoes unitary evolution. However, when disorder is sufficiently strong, the usual signatures of quantum chaos \cite{haake1991} disappear. Subsystems fail to thermalise and the entanglement entropy grows logarithmically in time, a phenomenon known as many-body localisation (MBL) \cite{oganesyan2007,znidaric2008,pal2010,bardarson2012,serbyn2013,huse2014,nandkishore2015,abanin2019,abanin2021,suntaj2020}.
Despite the universality of such behaviour, predicting and measuring entanglement dynamics remains notoriously difficult.

The signatures of chaotic and localised dynamics in many-body systems manifest across a broad range of physical quantities, which include level spacing ratio distributions \cite{oganesyan2007,atas2013}, the correlation hole of the spectral form factor (SFF) \cite{leviandier1986,brezin1997,delcampo2017,cotler2017}, the entanglement scaling of excited states, and the subsystem entanglement entropy of the experimentally accessible initial state. However, spectral statistics \cite{haake1991} requires precise knowledge of the many-body spectrum, which is nontrivial as exponential Hilbert space growth typically reduces level spacings below experimental resolution limits \cite{roushan2017,jirovec2025}. The SFF demands both the ability to randomise over initial states and to measure overlaps between initial and time-evolved states \cite{joshi2022,das2025,dong2025}. Verifying volume-law versus area-law entanglement scaling requires access to highly excited eigenstates, whereas experiments generally prepare simple product states in the computational basis. Our experiment therefore targets the subsystem entanglement entropy of an equal bipartition of a qubit chain, in particular the second R\'enyi entropy which requires only a measurement of subsystem purity, as obtained using our modified randomised measurement toolbox.

Below, we theoretically predict and experimentally verify on QuEra's Aquila quantum processor that the regimes of weak and strong disorder strength have qualitatively distinct patterns of entanglement growth. To this end, we measure the entanglement entropy as a function of evolution time and disorder in the tractable regime of six qubits and find entropy growth compatible with quantum chaotic (weak disorder) and localised (strong disorder) behaviour.

\begin{figure}
    \hspace{-0.3cm}
    \includegraphics[width=0.99\linewidth]{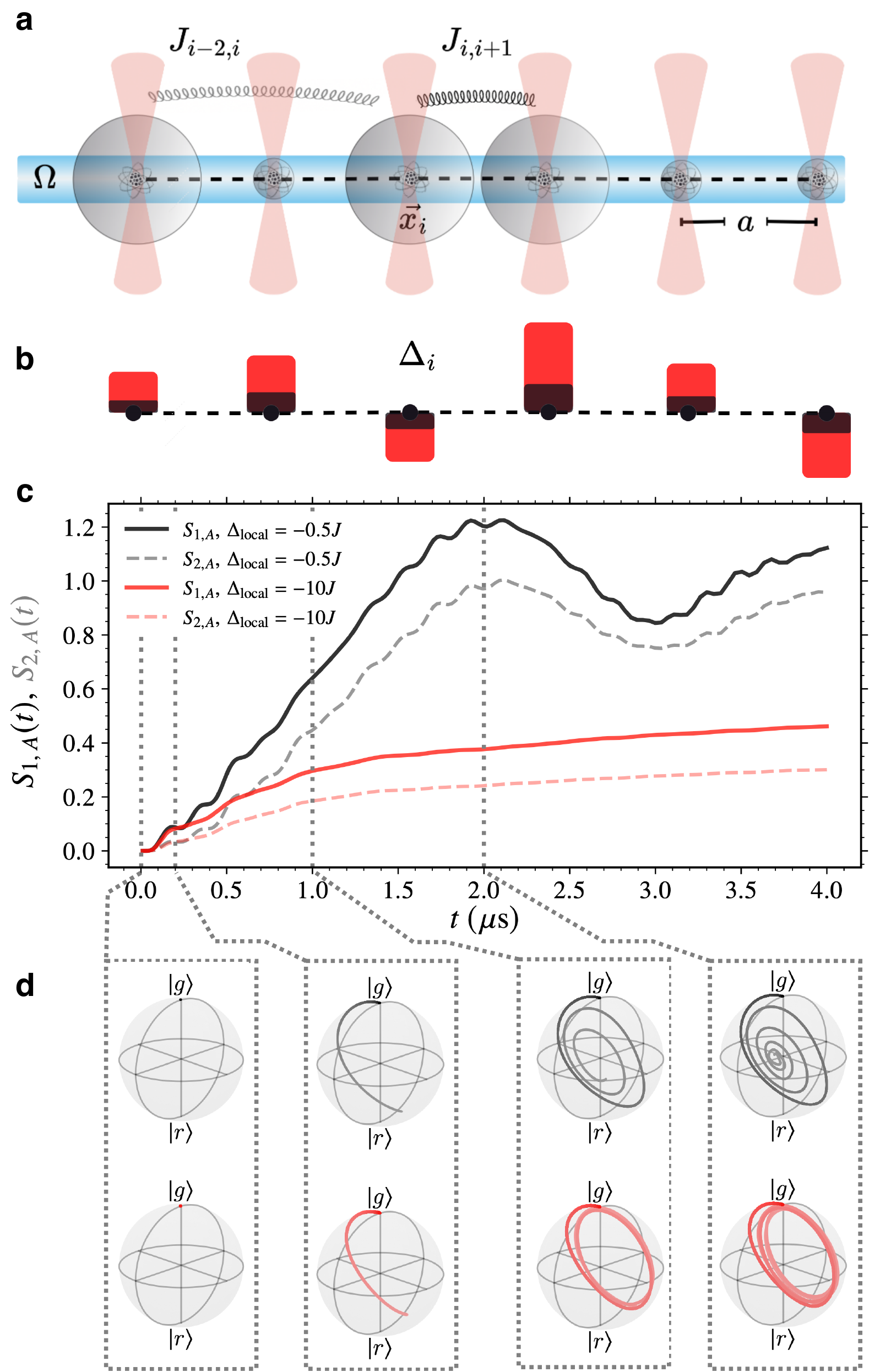}
    \caption{ \textbf{Disorder-dependent entanglement growth in a neutral-atom array.} {\sf{\textbf{a}}} Schematic of a chain of $^{87}\mathrm{Rb}$ atomic qubits trapped in optical tweezers with lattice constant $a$. Qubits are encoded in the $\ket{g}\equiv\ket{^5S_{1/2}}$ (small spheres) and $\ket{r}\equiv\ket{^{70}S_{1/2}}$ (large spheres) levels of $^{87}\mathrm{Rb}$ atoms. A global Rydberg drive (blue shade) excites the atoms to the Rydberg state and a Rydberg atom $i$ located at position $\vec{x}_i$ interacts with every other excited atom $j$ with interaction energy $J_{ij}$ due to the van der Waals interaction. 
    {\sf{\textbf{b}}} An illustration showing that each atom can be individually detuned from resonance with the global laser drive $\Omega$ by an amount $\Delta_i(t) = \Delta_{\mathrm{global}}(t) + \Delta_{\mathrm{local}}(t)h_i$ with a static coefficient $h_i$.  Here and in the rest of the figure, the nearest-neighbour strength is $J = 5.42 \mu \mathrm{s}^{-1}$, the global Rabi driving is $\Omega = 2.915J$, and we take two levels of disorder: weak, $|\Delta_{\mathrm{local}}| = 0.5J$ (black) and strong, $|\Delta_{\mathrm{local}}| = 10J$ (red). 
    {\sf{\textbf{c}}} Von Neumann entropy $S_{1,A}(t)$ (solid curves) and second-order R\'enyi entropy $S_{2,A}(t)$ (dashed curves) as a function of time for the subsystem $A$  composed of the first three qubits in a chain of $N=6$, averaged over $1000$ disorder realisations (see Appendix~\ref{chaoloc}).
    {\sf{\textbf{d}}} For each time slice in {\sf{c}} (grey dotted lines, $0.001 \mu \mathrm{s}, \, 0.2\mu \mathrm{s}, \, 1.0 \mu \mathrm{s}, \, 2.0 \mu \mathrm{s}$), the Bloch vector trajectory (from $t=0$) of the fourth qubit from the left in a chain of $N=6$ for a particular disorder realisation $h_i = [0.25,0.093,0.038,0.58,0.18,0.36]$.
    }\label{figure1}
\end{figure}

\section{Chaotic and localised dynamics from the many-body Hamiltonian}
\label{aquila}

We study the transition from chaos to localisation in a commercially available neutral-atom-based analog quantum computer, Aquila~\cite{wurtz2023}, which naturally embodies a long-range interacting transverse-field Ising model. The competition between power-law spin-spin couplings, transverse-field-induced quantum fluctuations, and disorder in this model system generates nonintegrable dynamics that enable fast information scrambling while still allowing the emergence of MBL as interactions are tuned~\cite{nandkishore2015,abanin2019,abanin2021}. Unitary evolution of the system, as illustrated in Fig.\,\ref{figure1}{\sf a}, is described by the time-dependent Hamiltonian 
\begin{align} \label{eqn:hamAquila}
    H(t)/\hbar&= 
    \frac{\Omega(t)}{2} \sum_{i=0}^{N-1}\left(e^{i\phi(t)}\hat\sigma_i^+ + e^{-i\phi(t)}\hat\sigma_i^-\right) \\ \nonumber
    &- \sum_{i=0}^{N-1} \Delta_{i}(t)\hat{n}_i 
    + \sum_{ i < j} J_{ij}\hat{n}_i\hat{n}_j , 
\end{align}
where $\hat{\sigma}_j^\pm= \hat{\sigma}_j^x \mp i \hat{\sigma}_j^y$, $\hat{n}_j = (\id - \hat{\sigma}_j^z)/2$ ($\hat{\sigma}_j^{x,y,z}$ are the Pauli matrices at site $j$), and $J_{ij} = C_6 / |\vec{x}_i- \vec{x}_j|^6$ ($C_6$ is the Rydberg interaction constant, see Methods). A global optical field drives all atoms between $ \ket{g}$ and $\ket{r}$ with a time-dependent Rabi frequency $\Omega(t)$, phase $\phi(t)$, and detuning $\Delta_\text{global}(t)$. A programmable local detuning $\Delta_i(t) = \Delta_{\mathrm{global}}(t) + \Delta_{\mathrm{local}}(t)h_i$ shifts the single-particle energy of each atom (Fig.\,\ref{figure1}{\sf b}), as controlled by static grayscale coefficients ${h_i}\in[0,1]$ ~\cite{cuadra2025}. For a lattice constant larger than the Rydberg blockade radius ($R_b=(C_6/\sqrt{\Omega^2+\Delta_i^2})^{1/6}$)~\cite{lukin2001,urban2009}, dipolar interactions lead to spread of entanglement throughout the qubit chain. Here we chose a lattice constant $a=10\,\mu\mathrm{m}$, and $\Omega(t)=15.8 \mu \mathrm{s}^{-1}$ to achieve a non-blockaded condition. 

In order to observe the transition from the localised to chaotic regimes, we programme the local disorder using the site-dependent local detuning parameters $\Delta_{\mathrm{local}}$ and $h_i$. A random configuration of the $h_i$ across the atoms ($h_i$ values drawn from a uniform distribution in $[0,1]$), for a fixed set of parameters $J_{ij}, \Omega, \Delta_{\mathrm{global}}$, and $\phi$ [Eq.~\eqref{eqn:hamAquila}], thus realises a distinct evolution under disorder.  
Within this parameter space, the chaotic regime is expected when $\Delta_i \lesssim \Omega \sim J$, while the localised regime is expected when $\Delta_i \gg \Omega \sim J$. 
Regime identification is confirmed by numerically evaluating spectral and dynamical signatures within the Aquila Hamiltonian (Eq.\,\eqref{eqn:hamAquila}), as fully detailed in Appendix~\ref{chaoloc}.

A conventional measure of entanglement is the von Neumann entropy, $S_{1,A}=-\mathrm{tr}\left(\rho_A \log \rho_A \right)$, which is a nonlinear (logarithmic) function of the eigenvalues of a subsystem's density matrix $\rho_A$; evaluating it generally requires access to the full spectrum, which typically entails full state tomography or comparably resource-intensive methods. We thus consider the second Rényi entropy $S_{2,A} = -\log\left(\operatorname{Tr}(\rho_A^2)\right)$, which is a natural quantity to measure experimentally with the subsystem purity $\operatorname{Tr}(\rho_A^2)$.
Fig.\,\ref{figure1}{\sf c} shows a numerical simulation of the von Neumann entropy (solid curves) and second Rényi entropy (dashed curves) as a function of time, starting from a polarised product state $\ket{g}^{\otimes N}$, for the low-disorder (black) and high-disorder (red) regimes. The initial growth is linear for the low-disorder case, while it is logarithmic for the high-disorder case, bearing the hallmarks of the chaotic and localised regimes, respectively. We also confirm that the two entropies exhibit the same qualitative dynamics across the transition (as has been shown for other Hamiltonians, e.g. the XXZ model \cite{fan2017}). The corresponding behaviour of the single-qubit Bloch vector trajectory and, importantly, its purity (given by the vector length) is shown in Fig.\,\ref{figure1}{\sf d} for the fourth qubit in a chain of $N=6$ for the low- and high-disorder cases at four time points. Initially, the Bloch vectors point up to $\ket{g}$; then they start to rotate on the surface. At later times, the Bloch vector for the chaotic realisation spirals in towards the centre, losing purity due to entanglement, whereas the high-disorder realisation remains close to the surface, maintaining purity.


\section{Randomised Measurements through Global Operations}

Randomised measurements \cite{vanenk2012,elben2019,elben2022} enable efficient extraction of information from many-body quantum systems by mapping correlations of the state onto measurement statistics generated by random unitary operations --- a method closely related to the technique of classical shadows \cite{huang2020}. In a typical randomised measurement protocol, one (i) prepares the state $\rho$, (ii) applies a randomly selected unitary $U_y$ from some ensemble to the subsystem of interest $\rho_A$, and (iii) measures the final state vector $\{ \ket{s}\}$
through projective measurement in the computational basis $n_S$ times. (i)-(iii) are repeated for $n_U$ different $U_y$, and one obtains estimates for the desired linear or nonlinear functionals of $\rho_A$ using classical postprocessing that exploits the statistics of the ensemble that $\{ U_y\}$ are sampled from. Modifications to this procedure applicable to systems where sampling unitary operations may be difficult include making use of disorder quenches \cite{elben2018, vermersch2018} or interaction quenches \cite{denzler2024}, and exploiting entanglement with ancilla systems \cite{mcginley2023, tran2023}.


Here we develop a novel randomised measurement toolbox that realises effective local rotations using only global controls. Since we want to make randomised measurements on subsystems, our scheme must be capable of implementing any unitary in $U(2^N)$. To achieve this, we revisit the early arguments of Lloyd~\cite{lloyd1995, lloyd1996} and  Deutsch, Barenco, and Ekert~\cite{deutsch1997}. 
Suppose the target $U_y$ has generator $Z_y$ (i.e., $e^{iZ_y} = U_y$), and that we have access to two non-commuting Hamiltonians $H_0$ and $H_1$.
By composing the evolutions $e^{i \tau_k H_0}$ and $e^{i \tau_k H_1}$ (over time $\tau_k$) in a suitable sequence, the nested commutators appearing in the Magnus expansion of the resulting propagator can be arranged to reproduce any element of the Lie algebra generated by $H_0$ and $H_1$.
If this algebra is the full $\mathfrak{su}(2^N)$, every generator $Z_y$---including all local and subsystem terms---is in principle accessible, and consequently any $U_y$ can be implemented.

\begin{figure}
    \centering
    \includegraphics[width=0.99\linewidth]{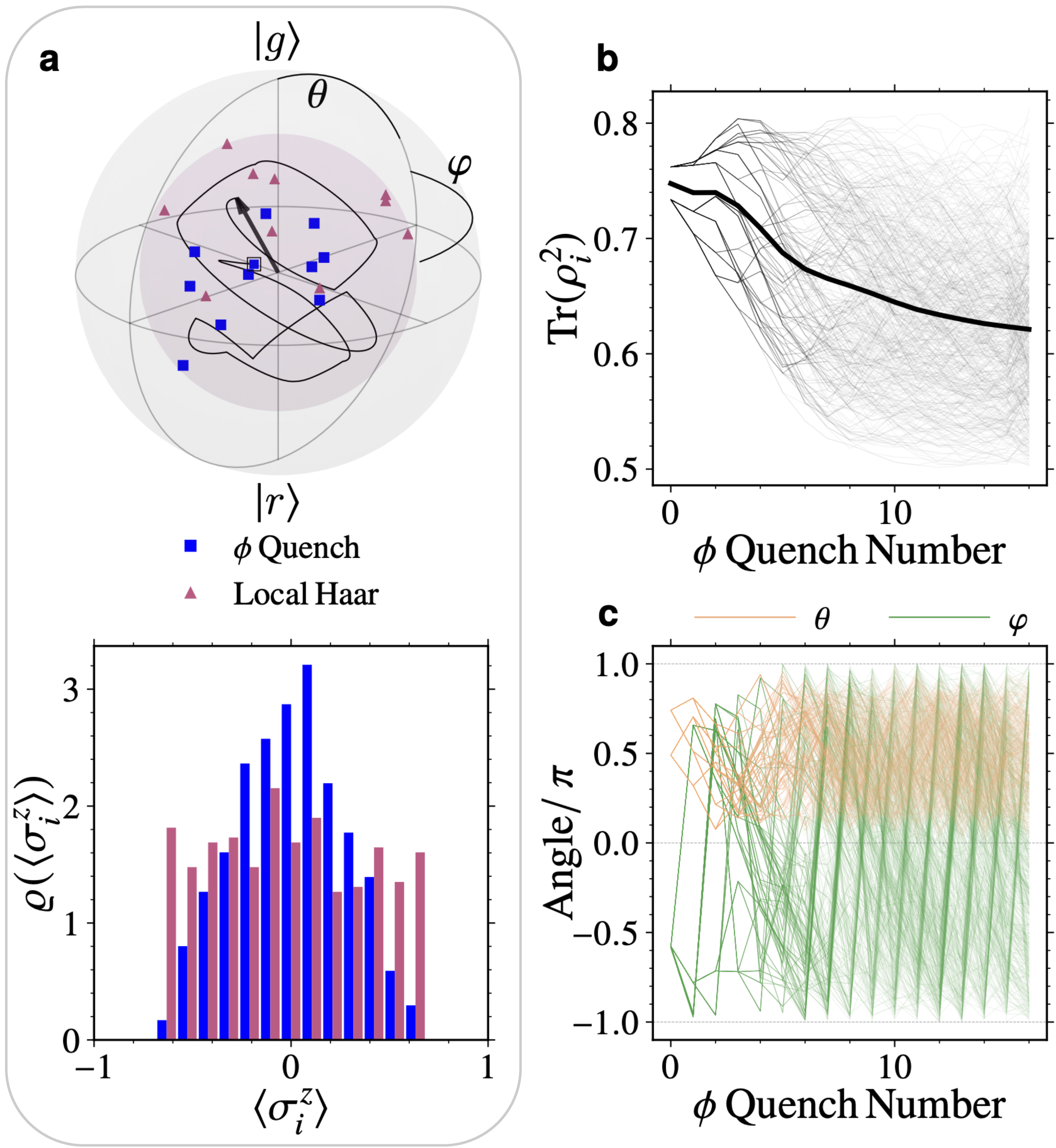}
    \caption{\textbf{Simulation of the modified randomised measurement protocol.} {\sf{\textbf{a}}} (Top) The black arrow is the Bloch vector of the third qubit in a chain of six atoms evolved for $1 \mu \mathrm{s}$ under a Hamiltonian drawn from the chaotic ensemble ($|\Delta_{\mathrm{local}}|=0.5J$); the inner purple sphere has radius of the length of the Bloch vector to be randomised, and the outer gray is the unit sphere. Blue squares are 10 realisations of our ``$\phi$-quench" sequence on the initial Bloch vector (black arrow). Rose triangles are single-qubit rotations, on the initial Bloch vector, drawn directly from the Haar distribution. The solid curve is the trajectory from the initial to the final vector after evolution under an example $\phi$-quench sequence. 
    (Bottom) The normalised discrete probability density (bin width of 0.0474) of $z$-axis projections of 500 Bloch vectors $\varrho(\langle \sigma_z^i\rangle)$ after applying $\phi$-quench gates (blue) and local Haar rotations (rose). 
    {\sf{\textbf{b}}} Purity after each additional $\phi$ quench ($0$ or $\pi/2$ with equal probability after a time $\tau = 0.06\mu \mathrm{s}^{-1}$ under $\Omega = 15.8\mu \mathrm{s}^{-1}$; solid black curve is the average across trajectories.
    {\sf{\textbf{c}}} Polar and azimuthal state angles on the Bloch sphere in {\sf a} as a function of $\phi$-quench number for individual $\phi$-quench sequences.
    }\label{figure2}
\end{figure}

As discussed above, the transition from chaotic to MBL behaviour is captured by the dynamics of the entanglement entropy (Appendix~\ref{chaoloc}), here reflected in the second Rényi entropy $S_{2,A} = -\log\left(\operatorname{Tr}(\rho_A^2)\right)$. Because $S_{2,A}$ is expressed in terms of purities of subsystems $\operatorname{Tr}(\rho_A^2)$, which is a second-order moment of the state, we only require that $U_y$ is sampled from a distribution that approximates a $2$-design \cite{elben2022}. The main difficulty lies in demonstrating that the Hamiltonians programmable in Aquila can generate, or sufficiently approximate, the required ensemble of subsystem rotations. For this, we create $H_0$ and $H_1$ from an implementation of the Aquila Hamiltonian (Eq.\,\eqref{eqn:hamAquila}) that is identical in all parameters except for the phase of the global drive, $\phi = 0$ ($H_0$) or $\phi = \pi/2$ ($H_1$); as discussed in Appendix\,\ref{apx_protocol}, this choice of $H_0$ and $H_1$ meets the universality condition set out above. When combined with randomly sampled local detunings $\Delta_i$, an ensemble of arbitrary $L$-step sequences of $\phi$ quenches effectively results in the ensemble of local single-qubit rotations needed for randomised measurements. As fully detailed in Appendix\,\ref{apx_protocol}, we perform a simple optimisation procedure to identify parameter regimes for randomisation at the level of single-qubit Bloch spheres.

Figure\,\ref{figure2}{\sf a} (top) shows the Bloch vector of the third qubit in a chain of six atoms after each realisation of a random $L=16$-step sequence (blue squares) starting from the same reference state (black arrow), and demonstrates coverage of the Bloch sphere comparable to a Haar-random state (rose triangles). Fig.\,\ref{figure2}{\sf a} (bottom) shows the corresponding distribution of $z$-basis measurements, and demonstrates that our protocol (blue) produces measurements with a variance that is $\approx 50\%$ of the Haar-random distribution. At the single-qubit level, the randomised measurement protocol works because the purity (length of the Bloch vector) is proportional to the spread in $\langle \sigma_i^z \rangle$ under uniform rotations.
The measurement results from our protocol are more peaked around $z=0$ due to a partial loss of purity. This is also shown in Fig.\,\ref{figure2}{\sf b}, where the qubit purity is tracked through the $16$ randomised $\phi$ quenches, showing a partial loss of purity on average, and for most trajectories. Coverage of the full Bloch sphere surface is also evidenced by tracking the polar and azimuthal angles, $\theta$ and $\varphi$, of the trajectories through the randomised sequence, as shown in Fig.\,\ref{figure2}{\sf c}. 

While this is only an illustrative example for a possible reference state, produced by a specific realisation of Eq.\,\eqref{eqn:hamAquila}, it is indicative that our protocol produces a sufficiently randomised state. As an ultimate verification, we find good agreement between a noiseless emulation of our randomised measurement protocol, yielding computational basis outputs, and the entropy calculated from numerically obtained reduced density matrices, as discussed in Appendix~\ref{apx_noiselss_protocol}. This demonstrates that the protocol produces close to the relevant unitary 2-design statistics.

\section{Measuring entanglement dynamics}

\begin{figure}
    \includegraphics[width=0.91\linewidth]{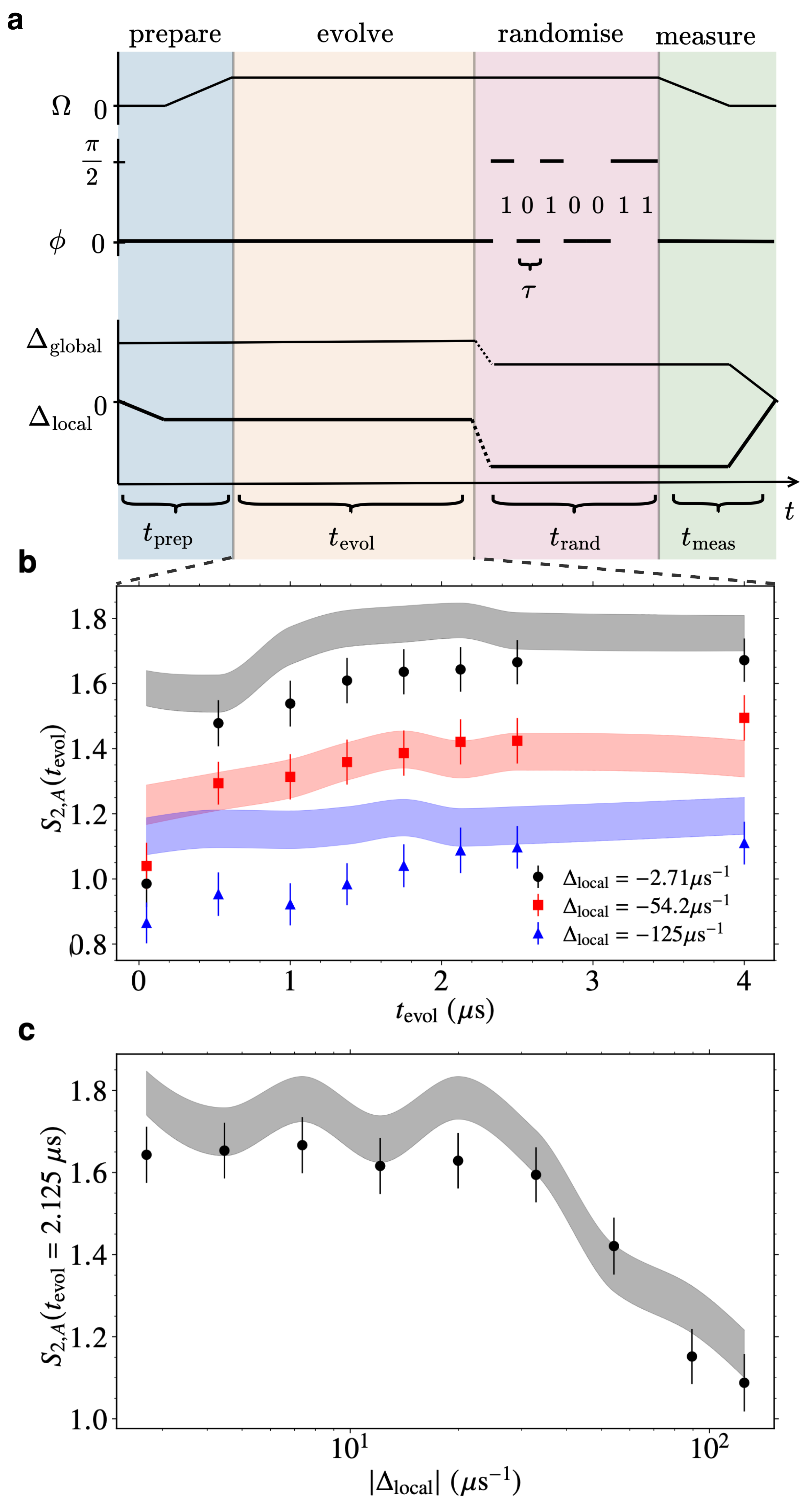}
    \caption{\textbf{Measurement of the disorder-dependence of entanglement entropy.} {\sf{\textbf{a}}} Experimental protocol showing the time dependence of the Aquila Hamiltonian parameters: $\Omega$ and $\Delta_{\mathrm{local}}$ are ramped to their target values; held for $t_{\mathrm{evol}}$; ramped to values optimised for randomised rotations (Appendix~\ref{apx_protocol}); held while $\phi \in \{0, \frac{\pi}{2}\}$ (shown as $\{0,1\}$) is chosen with $50\%$ probability $L=16$ times, each for a duration $\tau=0.06\mu\mathrm{s}$; and finally ramped to $0$ for projective measurement. We vary the width of the distribution of detunings per qubit but fix the mean to $2.71 \mu \mathrm{s}^{-1} = 0.5 J$ (Fig.\,\ref{figure1}{\sf b}).
    {\sf{\textbf{b}}} Experimentally measured entropy as a function of the evolution time $t_\mathrm{evol}$ for three values of disorder: $\Delta_{\mathrm{local}}= -0.5J$ (black circles), $\Delta_{\mathrm{local}}= -10J$ (red squares), $\Delta_{\mathrm{local}}= -23.06J$ (blue triangles). 
    {\sf{\textbf{c}}} Experimental entropy as a function of $|\Delta_{\mathrm{local}}|$ for a fixed evolution time $t_\mathrm{evol}= 2.125\mu \mathrm{s}$, i.e. when the entropy maximises in the noiseless $|\Delta_{\mathrm{local}}| = 0.5J$ case.  Error bars in {\sf b} and {\sf c} represent one standard error on the mean calculated from propagating multinomial shot noise. The shaded areas, shown in the respective color of each experiment in {\sf b} and {\sf c}, are from a numerical simulation of the protocol including the effects of readout error and decoherence. Their width represents one standard error on the mean evaluated from propagating multinomial shot noise (see Methods).}
    \label{fig:rand}
\end{figure}

We choose to use a chain of $N=6$ qubits to balance available system coherence with the entanglement dynamics we wish to see experimentally (see Methods). Figure\,\ref{fig:rand}{\sf a} outlines the complete experimental protocol tailored to Aquila. We begin with a preparation stage followed by an evolution stage where the system is evolved for a variable time ($t_\text{evol}$) under the time-independent Hamiltonian of Eq.~\eqref{eqn:hamAquila}. We tune $\Delta_i(t)$ to realise a chaotic or localised Hamiltonian. For both cases, the local coefficients $h_i$ are drawn uniformly in the range $[0,1]$ for each disorder realisation. After the evolution stage, we ramp up the local disorder $|\Delta_{\mathrm{local}}|$ to put the system in a localised regime where the single qubit purity is preserved (see Fig.\,\ref{figure1}{\sf d}). Then we utilise the phase control of Aquila (see Eq.~\eqref{eqn:hamAquila}) to engineer a random binary sequence, of length $L=16$, of phase $\phi$ quenches switching between $\phi=0$ (`$0$') and $\phi=\pi/2$ (`$1$'). After the randomisation sequence we perform projective measurement in the computational basis to obtain a single state vector $\{ \ket{s}\}$ (see Methods for more details).  

For each of unitaries $U_y$ that we apply in our randomised measurement sequence, we measure $n_{\mathrm{S}}$ shots and then take an average over $n_U$ to obtain an estimate of the probability of obtaining each computational basis element $\{ \ket{s}\}$ (in the randomised basis of the full system). From this we can construct the probabilities $P(s_A)$ of the computational basis elements corresponding to subsystem $A$, here consisting of the first three of $N=6$ qubits from left to right, after applying $U_y$ globally. The purity is then obtained via a weighted sum over all pairs of possible bitstrings for $A$ and averaged over the different applications of random unitary operations \cite{brydges2019}:
\begin{equation}
    \operatorname{Tr}(\rho_A^2) \approx 2^{N_A} \sum_{s_{A}, s'_{A}}(-2)^{-D[s_A, s'_A]}\overline{P(s_A) P(s'_A)},
    \label{eqn:RMTX_renyi}
\end{equation}
where $D[s_A, s'_A]$ is the Hamming distance (the number of places the two computational basis strings $s_{A}, s'_{A}$ disagree). From this we obtain the second-order R\'enyi entropy $S_{2,A} = -\log\left(\operatorname{Tr}(\rho_A^2)\right)$ for each Hamiltonian realisation. Finally, for each evolution time ($t_\text{evol}$), we average the entropy for $n_{\mathrm{ens}} = 15$ different disordered Hamiltonian realisations.

Figure\,\ref{fig:rand}{\sf b} shows the experimentally measured second-order R\'enyi entropy as a function of the duration of the evolution step for a low disorder regime $\Delta_{\mathrm{local}}=-0.5J$ (black squares), for a moderate disorder regime $\Delta_{\mathrm{local}}=-10J$ (red squares), and for the highest disorder reachable in Aquila $\Delta_{\mathrm{local}}=-23.06J$ (blue squares). The growth of entanglement entropy for the low-disorder case is faster and reaches a higher value than for the moderate-disorder case. We also observe that the entanglement in the low-disorder case reaches a peak, followed by a plateau, while in the moderate-disorder case it keeps growing slowly. The highest disorder case (blue) provides a simple reference: within error bars the entanglement entropy does not significantly grow from its $t_{\rm{evol}}=0$ value. Our data thus qualitatively agree with the disorder-dependent entanglement growth behaviour expected from our unitary simulation (Fig.\,\ref{figure1}{\sf c}). Numerical simulations that account for the effects of decoherence and readout errors in Aquila (see Methods) are shown, respectively, as the shaded black, red, and blue regions in Fig.\,\ref{fig:rand}{\sf b}, and are in good agreement with the data for most evolution times. In particular, these simulations explain the origin of the large time-independent entropy offset observed in all data relative to the fully unitary predictions of Fig.\,\ref{figure1}{\sf c}. 

Our data reveal clearly a statistically significant difference in entanglement growth between the three disorder regimes. Accordingly, Fig.\,\ref{fig:rand}{\sf c} shows the measured dependence on the disorder strength $|\Delta_{\mathrm{local}}|$ of the second-order R\'enyi entropy at a fixed evolution time $t_{\text{evol}} = 2.125 \mu \mathrm{s}$ (chosen to maximise dependence on disorder). The data again reveal a clear trend: at low disorder strengths ($|\Delta_{\mathrm{local}}|\lesssim 5J$), the system is presumed in a quantum chaotic phase, characterised by higher entropy values; as the disorder strength increases to moderate values ($|\Delta_{\mathrm{local}}|\sim 5-15J$), the system transitions into a ``nontrivial" localised phase, where the entropy decreases significantly -- but not altogether -- following a suppressed logarithmic growth with time, as seen quantitatively in our model (Appendix~\ref{chaoloc}). At the maximum disorder strength ($|\Delta_{\mathrm{local}}|> 15J$), the system is presumed in a ``trivial" localised phase: the dynamics are frozen out by the large average energy detuning of each qubit. These behaviours are consistent with the suppression of entanglement growth in the presence of significant disorder, as observed also in the numerical simulations that account for decoherence and readout errors (black shaded region). Taken together, the results of Fig.\,\ref{fig:rand}{\sf b,c} underscore the distinct dynamical regimes of the system shown more intuitively in Fig.\,\ref{figure1}{\sf d}. 
In the MBL phase, single qubits remain on the surface of the Bloch sphere (close to pure states), thus not building entanglement with their neighbouring qubits. 
In contrast, in the chaotic phase, they spiral toward the center of the Bloch sphere, losing purity quickly while becoming entangled with the rest of the system. 

\section{Conclusion}

The development of quantum computing devices relies on the precise control of entanglement dynamics in many-body quantum systems. Gaining such control involves two complementary aspects: first, one must understand how to experimentally tune the parameters that govern the system’s behavior, as described by theoretical models; second, one must accurately measure the outcomes of this tuning to verify theoretical predictions. In this work, we demonstrated control in both ways. 

Firstly, we engineered local disorder by tuning the Hamiltonian parameters of the system. Control over on-site disorder enabled us to switch, experimentally, between the slow entanglement growth characteristic of MBL and the rapid growth associated with quantum chaos. Verifying the existence of a phase transition will require a more extensive experimental campaign, including a scaling analysis with a large number of qubits. However, our experiment has established a baseline signature of the entanglement transition from the presumed MBL to the quantum chaotic regimes, as identified by the spectral and dynamical signatures numerically obtained for our system. From the data, we successfully estimated subsystem entropies and demonstrated how changing the Hamiltonian parameters allows us to tune the system across distinct entanglement-growth regimes. This study targeted one-dimensional arrays large enough to display emergent many-body phenomena yet still amenable to direct comparison between numerical simulations and current experimental capabilities. In particular, the experimental data matched theoretical predictions once false-detection error rates and decoherence were taken into account.


Secondly, on the measurement side of control, we proposed and implemented a protocol to measure second Rényi entropy and thus verify the distinct entanglement-growth behaviors that characterise the MBL and quantum-chaotic regimes. Specifically, we developed and applied a modified version of the randomised measurement toolbox, tailored for analog quantum processors without the need for local gate control, to experimentally obtain the entanglement entropy of arbitrary subsystems. Our protocol uses a set of unitary rotations that---despite being applied via a global control field---effectively approximate universal local operations. This paves the way for implementing randomised measurements in systems that lack local gate control, which is broadly applicable and requires fewer control resources. Moreover, we note that our method is directly applicable for measuring a range of physical quantities that include the SFF~\cite{leviandier1986,delcampo2017,joshi2022,das2025,dong2025}.

More broadly, our approach to tackling scalability and decoherence is to first control entanglement growth and many-body dynamics in the smallest systems that already exhibit the essential phenomena, and only then to extend these insights to larger architectures. While disorder is usually seen as a nuisance to be avoided, commercially available Rydberg atom arrays \cite{wurtz2023} offer precise control over it by engineering the local Hamiltonian parameters. This could elucidate the relationship between disorder and inhomogeneous decoherence with increasing system size \cite{kropf2016}. By dynamically modulating the disorder in real time, one could steer the system back and forth between localised and chaotic phases, thereby providing a direct handle on the growth of many-body entanglement.

\acknowledgments
We thank David Huse for stimulating discussions about MBL and our theoretical results. We thank Peter Zoller and Mikhail D. Lukin for the lecture series  (Quantum Connections 2023, Sweden) and discussions on randomised measurements and Rydberg simulators that formed the conceptual foundation of this project. We acknowledge support from Amazon Web Services (AWS) through the provision of Amazon Braket credits used for quantum computing experiments on QuEra Aquila. This project was in part funded and supported by the UK National Quantum Computer Centre [NQCC200921], which is a UKRI Centre and part of the UK National Quantum Technologies Programme (NQTP). A.S.M.-R. was supported by the EPSRC UK Multidisciplinary Centre for Neuromorphic Computing (grant UKRI982). L.~S.~was supported by a Research Fellowship from the Royal Commission for the Exhibition of 1851. A.~D.~thanks support from the EPSRC through the QQQS programme grant (EP/Y01510X/1) and the QCi3 hub (EP/Z53318X/1). D.A.G acknowledges a Royal Society University Research Fellowship. 

\section*{Data Availability}
All raw data from AWS Braket and for our numerical simulations are publicly accessible via Oxford University Research Archive \cite{scholin2026}.
All code to produce all of the figures and results of this work is publicly available on GitHub: \url{https://github.com/oscars47/RydSFF}.

\section{Methods}

\subsection{Experimental apparatus and protocols}
The experiments reported in this work were performed in the programmable analog quantum simulator Aquila developed and operated by QuEra Computing Inc. Aquila is based on $^{87}$Rb atom arrays trapped in optical tweezers that are driven to a highly excited Rydberg state, to create van der Waals interaction between them. The Rydberg interaction is static and can be programmed by specifying the positions of up to $256$ atoms within a two-dimensional area of $75\mu m$ by $125\mu m$. The amplitude, frequency and phase of the laser drive are fully programmable, enabling control over the corresponding time-dependent Hamiltonian parameters: Rabi frequency $\Omega(t)$, global laser detuning from ground-Rydberg resonance $\Delta_{\mathrm{global}}(t)$, and phase $\phi(t)$ (see Eq.~\eqref{eqn:hamAquila}). Local component of the detuning $\Delta_{\mathrm{local}}(t)h_i$ is controlled by an additional array of off-resonant laser beams, that create local AC Stark shifts of the ground-Rydberg transition relative to the global laser drive. The static spatial pattern and static intensities of the second array is controlled via the coefficients $h_i\in [0, 1]$. A detailed description of the experimental system and capabilities can be found in ~\cite{wurtz2023, cuadra2025}. In this section, we describe the relevant experimental protocols and sources of error. 

\subsubsection{Experimental protocol}
In this experiment we employ a system of $N=6$ atoms arranged in a 1D lattice with interatomic separation of $a=10\mu m$. The Hamiltonian of the system is given by Eq.~(\ref{eqn:hamAquila}).

Fig.\,\ref{fig:rand}{\sf a} outlines the experimental protocol tailored to Aquila, which comprises four stages: preparation, evolution, randomisation, and measurement.

\begin{enumerate}[(i)]
    \item \textit{Preparation}: $\Omega$ and $\Delta_{\mathrm{local}}$ are ramped up from zero to their target values at their maximum rates (see Appendix \ref{apx_experiment}), as described in the \textit{Evolution} phase. $\phi$ is fixed to $0$.
    \item \textit{Evolution}: The system is evolved under the programmed time-independent Hamiltonian of Eq.~\eqref{eqn:hamAquila} for a variable duration $t_\text{plateau}$. Setting $\langle \Delta_i \rangle = \Delta_{\mathrm{global}} + \Delta_{\mathrm{local}}/2$ makes $\abs{\Delta_{\mathrm{local}}}$ the disorder strength, as it determines the spread of detunings around the mean $\langle \Delta_i \rangle$. Each Hamiltonian corresponds to a random selection of $h_i$ values, and we tune the disorder strength to either $\Delta_{\mathrm{local}}=-0.5J$ (low-disorder chaotic phase), $\Delta_{\mathrm{local}}=-10J$ (high-disorder localised phase), or $\Delta_{\mathrm{local}}=-23.06J$ (trivially localised phase). To maintain $\langle \Delta_i \rangle = 0.5J$, we set the global detuning to $\Delta_{\mathrm{global}} = 0.75J$ (low-disorder chaotic phase), $\Delta_{\mathrm{global}} = 5.5J$ (high-disorder localised phase), or   $\Delta_{\mathrm{global}} = 12.03J$ (trivially localised phase).    
    For all experiments, we sample $15$ disorder realisations. 
    \item \textit{Unitary randomisation}: After the evolution $|\Delta_{\mathrm{local}}|$ is ramped up to about 80\% of its maximum value, $\Delta_{\mathrm{local}}=-102.7\mu \mathrm{s}^{-1}$, to ensure the system is in a localised regime ($\Delta_i \gg \Omega \sim J$) that approximately preserves the single-qubit purity, see Fig.\,\ref{figure1}{\sf d}. $\Omega$ and $\Delta_i$ are held fixed while we implement the rotations via a sequence of random $\phi$-quenches: at each step, we apply either $H_0$ ($\phi = 0$) or $H_1$ ($\phi = \pi/2$) for a duration $\tau=0.06\mu\mathrm{s}$. The switching pattern is generated by sampling a random binary string $s \in \{0,1\}^L$, where $L=16$ and $s_k=0$ selects $H_0$ and $s_k=1$ selects $H_1$. The total randomisation time is therefore $L\tau = 0.96\mu\mathrm{s}$. For every $t_{\mathrm{evol}}$ value of interest, we repeat this with $n_U$ different random rotations (when the evolution phase has $\Delta_{\mathrm{local}}=-0.5J$, $n_U = 15$; when  $\Delta_{\mathrm{local}}=-10J$, $n_U = 20$; when  $\Delta_{\mathrm{local}}=-23.06J$ , $n_U = 25$).
    When probing the disorder dependence in Fig.\,\ref{fig:rand}{\sf c}, the first data point at $\Delta_{\mathrm{local}}=-0.5J$ has $n_U = 15$; the next seven data points to the right have $n_U = 20$; the final point at $\Delta_{\mathrm{local}}=-23.06J$ has $n_U = 25$. 
    The higher $\Delta_i$ cases require more rotations because the broader spread of parameters at larger disorder strengths leads to markedly different dynamics across the ensemble (Appendix~\ref{chaoloc}). 
    The values of $L$, $\tau$, $\Delta_{\mathrm{local}}$, and $\Delta_{\mathrm{global}}$ are determined by numerical optimisation (see Appendix~\ref{apx_protocol}), while the number of rotations and shots are limited by resources.

    After the completion of this work we have seen a preprint of a related work by Toga et al.~\cite{toga2026}, which implements a global $2$-design on Aquila to measure out-of-time-order correlators. By contrast, our modified measurement toolbox measures the purity of subsystems, which inherently requires (approximately) unitary rotations on the subsystems of interest.
    
    \item \textit{Measurement}: we ramp $\Omega$,  $\Delta_{\mathrm{local}}$, and $\Delta_{\mathrm{global}}$ back to zero and perform a projective measurement in the computational basis; each shot yields a bitstring for each such projective readout. For each configuration of the Hamiltonian, we take 200 shots. We postselect the final bit-strings on shots that had a perfectly filled initial lattice and obtain on average 95\% correctly filled six-qubit arrays.
\end{enumerate}

The total number of shots for data presented in Fig.\,\ref{fig:rand}b,c is $1,800,000$.

\subsubsection{Ramps}

Since the hardware constraints require $\Omega$ and $\Delta_{\mathrm{local}}$ to be $0$ at the start and end of each sequence, we apply linear ramps of $\Omega$ and $\Delta_{\mathrm{local}}$ in the preparation and measurement steps (see Fig.\,\ref{fig:ramps_annotated} in Appendix \ref{apx_experiment}). We chose the shortest allowed ramp durations of $0.632 \mu\mathrm{s}$ for $\Omega$ and $0.05\mu\mathrm{s}$ for $\Delta_{\mathrm{local}}$. We also find that ramping up $\Omega$ after  $\Delta_{\mathrm{local}}$ in the \textit{Preparation} stage and similarly ramping down $\Omega$ before  $\Delta_{\mathrm{local}}$ in the \textit{Measurement} stage greatly improves the agreement between theoretical and experimental results for short time evolutions ($< 0.2 \mu \mathrm{s}$) as illustrated in Table \ref{tab:phase_vals} and Fig.\,\ref{fig:ramps_annotated} in Appendix \ref{apx_experiment}. 

\subsection{Experimental imperfection and error modelling}

\subsubsection{Readout error}
Due to experimental imperfections such as atom loss and imperfect optical pumping, there is a finite readout error in both ground and Rydberg states, parametrised by $\epsilon_g$ and $\epsilon_r$, respectively. To account for the non-negligible readout error during measurement, we define a map $T$ from the pre-measurement probabilities $(p_g,p_r)$ to actual measured probabilities $(\tilde{p}_g,\tilde{p}_r)$:
\begin{align}
    \begin{bmatrix}\label{eqn:correction}
        \tilde p_g \\ \tilde p_r
    \end{bmatrix} &=  T \begin{bmatrix}
        p_g \\  p_r
    \end{bmatrix},\\
    T &= \begin{bmatrix}
        p(g|g) & p(g|r) \\ p(r|g) & p(r|r)
    \end{bmatrix} = 
    \begin{bmatrix}
        1-\epsilon_g & \epsilon_r \\ \epsilon_g & 1-\epsilon_r
    \end{bmatrix},
\end{align}
where the notation $p(a|b)$ means the probability of measuring $a$ given that $b$ is the ``correct'' outcome. For an $N$ qubit probability vector $\mathbf{p}$, we apply $T$ to every qubit, hence the total transformation is:
\begin{equation}\label{eqn:Ntrans}
    \tilde{\mathbf{p}}  = T^{\otimes N} \; \mathbf{p}.
\end{equation}
We estimate $\epsilon_r=0.1$ and $\epsilon_g=0.05$ by performing single-qubit Rabi oscillations prior to running the task for each Hamiltonian in the ensemble, as we describe in Appendix \ref{sec:readout_calibrate}.

\subsubsection{Decoherence error}
Application of local detuning ($\Delta_{\mathrm{local}}$) introduces significant decoherence. In order to benchmark this, we perform a Rydberg-Ramsey measurement in the presence of applied uniform local detunings on each site of a $5\times4$ square site array spaced by $21\mu m$. We extract $T_2^*$ values by fitting a gaussian decay of the ramsey fringes. To be more specific, we vary the magnitude applied local detuning to obtain the $T_{2,\mathrm{bare}}^*$ with no local detunings applied, and the local-detuning-induced decoherence metric $\theta_\mathrm{ramsey}$ defined as the phase angle accumulated by evolution under the given local detuning within the local-detuning-induced decoherence time $\theta_\mathrm{ramsey}\equiv \Delta_{\mathrm{local}}\times T^*_{2,\mathrm{LD}}$~\cite{cuadra2025}. The decay is modeled using the equation: 
\begin{align}\label{eqn:T2star}
    \frac{1}{T_2^*} &= \frac{1}{T_{2, \mathrm{bare}}^*} + \frac{\Delta_{\mathrm{local}}}{\theta_{\mathrm{ramsey}}},
\end{align}
and using the latest benchmarking data from the experiment, we find  $T_{2, \mathrm{bare}}^* = 6.2 \pm 0.3 \mu \mathrm{s}, \, \theta_{\mathrm{ramsey}} = 4.8 \pm 0.4$, with $\Delta_{\mathrm{local}}$ in $\mu \mathrm{s}^{-1}$.

\subsection{Choice of $N$}

We chose to run our experimental protocol on $N=6$ qubits for three reasons. First, it is the largest system size for which we can resolve both the growth and the beginning of the saturation of the entanglement entropy within the experimentally available time window of $t_{\mathrm{evol}}\leq 3\mu\mathrm{s}$ (the coherence time is $4\mu\mathrm{s}$ and we need $1\mu\mathrm{s}$ for preparation, randomisation, and measurement). Second, the experimental error is dominated by single-qubit readout errors, which accumulate exponentially with system size; hence, we seek the smallest system size for which we can clearly distinguish the two regimes. Finally, random matrix spectral correlations, a hallmark of quantum chaos, become pronounced for Hilbert space dimensions $d \gtrsim 50$, requiring at least $N = 6$ (i.e., $d = 2^6$) \cite{haake1991}.

\subsection{Numerics and Experimental Data Processing}

 To obtain the error bars on the experimental data in Fig.\,\ref{fig:rand}{\sf a,b}, we  include multinomial shot noise with covariance $(\mathrm{diag(\mathbf{p})} - \mathbf{p}\mathbf{p}^T)/n_{\mathrm{shots}}$ where $\mathbf{p}$ is the experimental probability vector and $n_{\mathrm{shots}} \leq 200$ is the number of experimental shots. We postselect on only fully populated initial arrays which reduces the useable number of shots to around 190 on average.
This is propagated through the calculation of $\overline{P(s_A) P(s'_A)}$ and hence the purity for an individual Hamiltonian and through to the whole ensemble using \texttt{uncertainty.unumpy} (version 3.2.2) and adding the standard errors in quadrature. 

The shaded areas in Fig.\,\ref{fig:rand}{\sf a,b} simulate the experimental protocol but address readout error and decoherence error.  We use QuTip (version 5.1.0) in Python 3.12.2 to define the Aquila Hamiltonian in Eq.~\eqref{eqn:hamAquila}. We evolve an $N=6$ qubit state vector initialised at $|gggggg\rangle$ using \texttt{qutip.sesolve} with the same time-dependent structure of the Aquila Hamiltonian as the experiment but with the following modifications to account for inhomogeneous decoherence.
We resample the values of the local detuning for the \textit{Evolution} and \textit{Randomisation} steps for each qubit separately from a Gaussian centered at the original $\Delta_{\mathrm{local}}$ value with uncertainty $\sqrt{2}/T_2^*(h_i\Delta_{\mathrm{local}})$ for each single shot of the experiment emulated in QuTip with unitary evolution, where the dependence of $T_2^*$ on $\Delta$ is given by Eq.~\eqref{eqn:T2star}.
We compile all 200 shots per time per disorder realisation per rotation and apply Eq.~\eqref{eqn:correction} to obtain the readout-error corrected theoretical probability vector $\tilde{\mathbf{p}}$, which we then postprocess in the same way as for the experimental data. To smoothly connect the emulated points to create the shaded area, we use \texttt{scipy.interpolate.
PchipInterpolator} (version 1.15.2) which implements a piecewise cubic Hermite interpolating polynomial fit in $t_{\mathrm{evol}}$ and $\log_{10}(|\Delta_{\mathrm{local}}|)$  for Fig \ref{fig:rand}{\sf b,c} respectively and plotted the envelope as $\hat{\mu}(x)\pm\hat{\sigma}(x)$.

\bibliography{RydSFF}
\clearpage

\appendix

\section{Transition from Quantum Chaos to Localisation}\label{chaoloc}

Generic (that is, not fine-tuned) systems are usually found in a so-called quantum chaotic regime, which can be diagnosed through spectral statistics \cite{haake1991}. According to the Bohigas-Giannoni-Schmit (BGS) conjecture \cite{bohigas1984}, quantum systems whose classical counterparts are chaotic exhibit Wigner–Dyson \cite{dyson1962} level spacing distributions characteristic of random matrix theory. Quantum chaotic systems usually thermalise -- known as quantum ergodicity in the Eigenstate Thermalisation Hypothesis (ETH) community. As a consequence, excited energy eigenstates are expected to obey a volume law where the entanglement entropy of a subsystem is proportional to its size \cite{abanin2019,hastings2007}. Moreover, the entanglement entropy of initial states grows linearly in time.

In the localised phase, the spectrum is uncorrelated and thus exhibits an exponential level spacing distribution. At the phenomenological level, initial product states exhibit only logarithmic entanglement growth in time, and the Hamiltonian’s excited eigenstates obey an area law, with the entropy of a subsystem scaling with the size of its boundary. 

To study the transition from a many-body quantum chaotic regime to MBL, we tune the degree of disorder in the Hamiltonian parameters [Eq.~\eqref{eqn:hamAquila}]. We focus on onsite disorder introduced through a uniform distribution of local detuning coefficients $\Delta_i$ shown schematically in Fig.\,\ref{figure1}{\sf b} of the main text, and examine how its strength, $\abs{\Delta_{\mathrm{local}}}$, affects the transition. Instead of examining a single Hamiltonian realisation, whose dynamics is highly irregular for small systems (for which we cannot invoke self-averaging), one considers an ensemble of similar systems and averages observables over it. Hamiltonian averaging smooths out sample-specific irregularities and highlights universal spectral and dynamical properties.

\begin{figure}
    \hspace{-0.5cm}
    \includegraphics[width=1.04\linewidth]{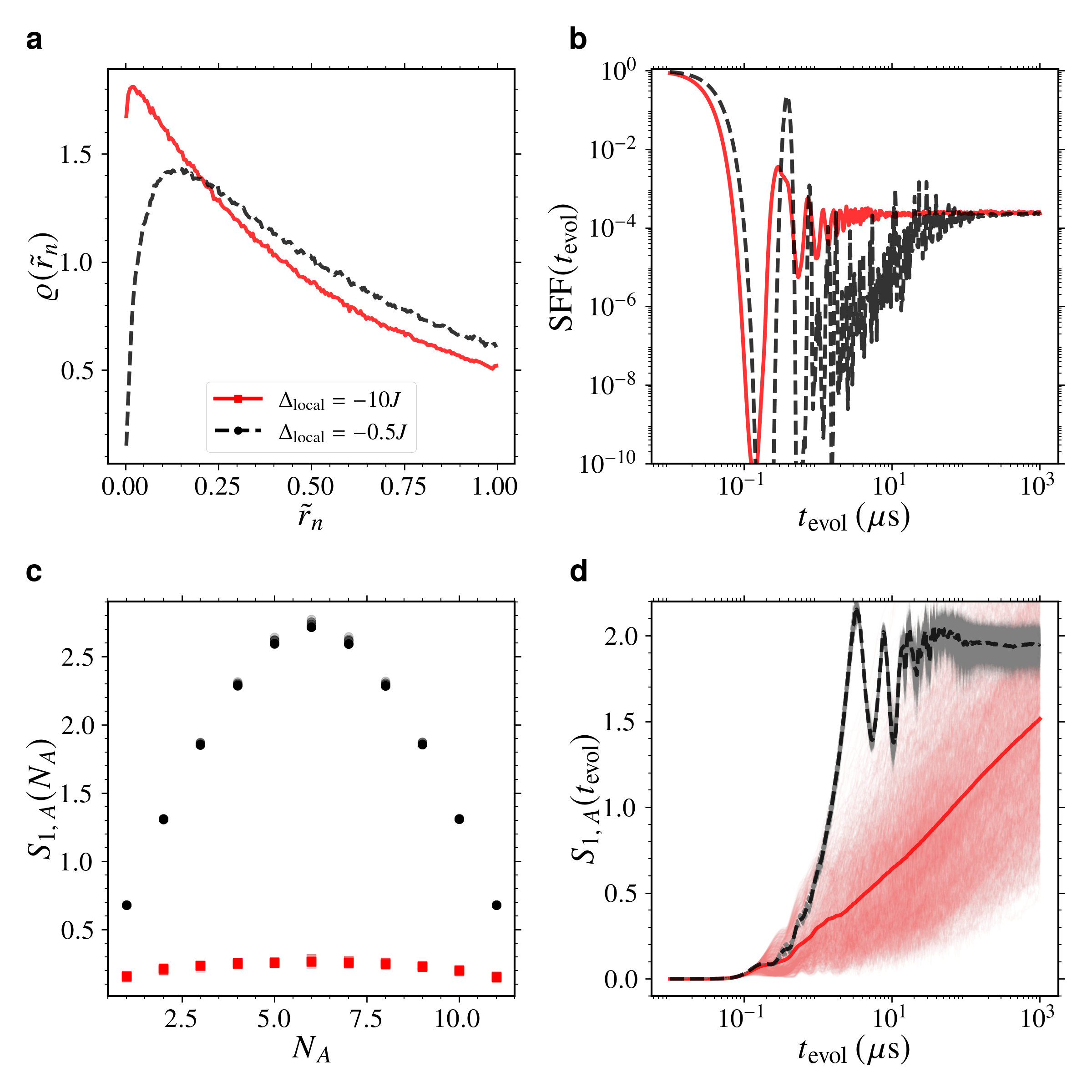}
    \caption{Spectral and dynamical signatures of the disorder-induced entanglement transition for the Aquila Hamiltonian [Eq.~\eqref{eqn:hamAquila}] with time-independent parameters. In all cases, we take $N=12$ qubits with 1000 disordered realisations, and $\langle \Delta_i \rangle = 0.5J$, $J = 5.42 \mu \mathrm{s}^{-1}$ nearest neighbor interaction strength, and Rabi drive $\Omega = 2.92 J$. Black curves and circles correspond to $\Delta_{\mathrm{local}}=-0.5J$, and red curves and squares to $\Delta_{\mathrm{local}}=-10J$.
    {\sf{\textbf{a}}} Probability density for the level spacing ratios $\tilde{r}_n=\min\{r_n,1/r_n\}$, where $r_n = s_n/s_{n-1}$ ($s_n=E_{n+1}-E_n$) To produce the histogram, the data is binned uniformly into 200 bins.
    {\sf{\textbf{b}}} Spectral form factor as a function of evolution time.  
    {\sf{\textbf{c}}} von Neumann entropy of ten excited states around the middle of the spectrum, where lighter markers corresponding to higher excited states, as a function of subsystem size. The subsystem of interest $A$ is taken to be the first $N_A$ qubits of the chain.
    {\sf{\textbf{d}}} von Neumann entropy, of the evolved $|g\rangle^{\otimes N}$ state as a function of evolution time, fixing $N_A = \lfloor N/2\rfloor$. The average is shown as a solid curve; individual realisations are shown as faint lines.
    }\label{signatures}
\end{figure}

We show numerically in Fig.\,\ref{signatures} that averaging over disordered realisations of the Aquila Hamiltonian [Eq.~\eqref{eqn:hamAquila}] yields distinctly different spectral and dynamical signatures, obtained by exact diagonalisation, depending on the value of $\abs{\Delta_{\mathrm{local}}}$, as predicted by the theories of many-body quantum chaos, quantum ergodicity, and its breaking. 
The system is in the chaotic phase when the disorder and interaction strength are comparable, $\abs{\Delta_{\mathrm{local}}} \sim  J$, whereas it is localised when $\Delta_{\mathrm{local}}\gg J$. We choose $\Delta_{\mathrm{local}}=-0.5J$ and $\Delta_{\mathrm{local}}=-10J$ as representative values of each case. We fix $\langle \Delta_i \rangle=0.5J$, where $J=C_6/a^6\approx 5.42\mu \mathrm{s}^{-1}$ is the average interaction energy between nearest neighbours for lattice constant $a = 10 \mathrm{\mu} m$, which guarantees that interactions remain relevant and the atoms do not simply polarise due to the external field (see Appendix \ref{ap:magnetisation}).
Since the disorder strength sets the maximum allowed detuning, and in order to retain the freedom to explore small $\abs{\Delta_{\mathrm{local}}}$, we fix the Rabi frequency at its maximal value of $\Omega = 15.8 \mu$s$^{-1}$ to minimise the blockade radius and ensure that every atom always lies outside the blockade radius of its neighbours. Throughout the evolution we set $\phi=0$.
In this numerical study, we chose to work with $N=12$ qubits for two reasons.
First, this is the largest equally spaced chain parallel to the experimental grid that allows one to probe the detuning regimes of interest without inducing blockade effects.
Second, given the minimum achievable spacing and maximum $\Omega$, this is the the largest system for which the half-partition entropy attains its maximum within the experimentally available coherence time ($4 \mathrm{\mu}$s).

\begin{table*}       
  \centering
  \begin{tabular}{|c|c|c|c|c|}
    \hline
     & Level spacings & SFF & Eigenstate  EE  & Entanglement growth\\ \hline
    Quantum chaotic & Wigner-Dyson & correlation hole & volume law & \makecell{linear growth,\\fast saturation} \\   \hline
    MBL & exponential & no correlation hole & area law & logarithmic growth \\ \hline
    Figure & Fig.\,\ref{signatures}{\sf a} & Fig.\,\ref{signatures}{\sf b} & Fig.\,\ref{signatures}{\sf c} &  \makecell{Fig.\,\ref{figure1}{\sf c};\;  Fig.\,\ref{fig:rand}~{\sf b,c}\\  Fig.\,\ref{signatures}{\textbf{d}};\; Fig.\,\ref{fig:bloqade_sim}~{\sf a,b}} \\ \hline
  \end{tabular}
  \caption{Signatures of many-body quantum chaos and MBL.}
  \label{tab:chaos_sig}
\end{table*}

\subsection{Numerical signatures of quantum chaos and localisation}

Under these conditions, we first look at the eigenvalue statistics of the Hamiltonian, as measured by the distribution $\varrho(r)$ of ratios of differences between adjacent eigenvalues, $r_n=s_n/s_{n-1}$, with $s_n=E_{n+1}-E_n$ and $E_n$ the ordered eigenvalues of $H$. In this context, quantum chaotic systems exhibit level repulsion in their level spacing distributions, meaning degeneracies are avoided, and energy eigenvalues are strongly correlated. On the other hand, MBL systems display level clustering and energy eigenvalues become effectively uncorrelated. Fig.\,\ref{signatures}{\sf a} confirms this behaviour for our parameter choices: for $\Delta_{\mathrm{local}}=-0.5J$ (chaotic phase), $\varrho(r)$ exhibits a peak and reflects weak eigenvalue correlations, while for $\Delta_{\mathrm{local}}=-10J$ (localised regime), $\varrho(r)$ is exponential and reflects strong eigenvalue correlations.
 
A corresponding dynamical characterisation of these regimes is provided by the SFF defined as $\mathrm{SFF}(t) = \overline{ |\mathrm{tr}\exp\{-i H t\}|^2} / 4^N$, 
where the average over Hamiltonians is denoted with an overbar, where we sample each $h_i$ randomly from a uniform distribution from 0 to 1 for each qubit for each disorder realisation. We compute the SFF from the exact eigenvalues computed before; it can alternatively be defined as the Fourier transform of the two-point correlation function of the energy spectrum \cite{leviandier1986}, the survival probability of the infinite-temperature coherent Gibbs state \cite{delcampo2017}, or the square of the absolute value of the normalised analytically continued partition function \cite{cotler2017}. 
A quantum chaotic system is characterised by a correlation hole in its SFF---a characteristic dip-ramp-plateau structure---while a regular Hamiltonian does not show such a feature. 
Fig.\,\ref{signatures}{\sf b} confirms this behaviour for our parameter choices. In both regimes, there is an initial oscillatory decay due to the non-homogeneity of the spectral density and a saturation to a plateau at long times. 
For weak disorder ($\Delta_{\mathrm{local}}=-0.5J$), the SFF dips below the plateau (i.e., it displays a correlation hole) and then shows a clear linear ramp. In contrast, for strong disorder ($\Delta_{\mathrm{local}}=-10J$), the SFF saturates directly to the plateau. 

The disorder-induced transition also manifests in the entanglement content of subsystems. In a one-dimensional system, the eigenstates in the MBL phase are expected to have negligible entanglement between spatially separate subsystems, while in the chaotic phase, the entanglement entropy of excited eigenstates is expected to grow with subsystem size. As a measure of subsystem entanglement, we compute the von Neumann entropy, $S_{1,A}=-\mathrm{tr}\left(\rho_A \log \rho_A \right)$, where $\rho_A=\mathrm{tr}_{A'}|n\rangle\langle n|$ is the reduced density matrix of an excited eigenstate $\ket{n}$ in a subsystem $A$ obtained by tracing over its complement $A'$. 
In Fig.\,\ref{signatures}{\sf c}, we compute the subsystem $A$ von Neumann entropy scaling using the ten excited states of the Hamiltonian that are closest to the experimentally accessible state $|g\rangle^{\otimes N}$ which for our parameter regimes is the middle of the spectrum.
Starting with the leftmost qubit alone, $N_A=1$, we add one qubit at a time, construct the reduced density matrix, and perform a singular‑value decomposition to obtain the Schmidt coefficients $\lambda_i$. The subsystem entropy is then $S_{1,A} = \sum_i \lambda_i \ln \lambda_i$.
At low disorder strength ($\Delta_{\mathrm{local}}=-0.5J$) the entropy grows with the subsystem size $N_A$ (volume law), whereas at high disorder strength ($\Delta_{\mathrm{local}}=-10J$) it saturates to an approximately constant value (area law). Further evidence for the eigenstate properties of the two regimes is show in Appendix~\ref{app:S_eigen}.

As shown in Fig.\,\ref{signatures}{\sf d}, the time evolution of the entanglement entropy provides a sharp dynamical distinction between the chaotic and MBL regimes. 
Starting from a product state $|g\rangle^{\otimes N}$ around the middle of the spectrum, we evolve in the energy‑eigenbasis. 
At each time step, we produce the reduced density matrix. 
A singular‑value decomposition of this matrix gives the time‑dependent singular values $\lambda_i (t)$, from which the von Neumann entropy is obtained as $S_{1,A}(t)= \sum_i \lambda_i(t) \ln \lambda_i(t)$. For weak disorder, $S_{1,A}(t)$ exhibits rapid growth, followed by a transient overshoot, and relaxation to a thermal plateau consistent with ETH and volume-law scaling. 
By contrast, at strong disorder, $S_{1,A}(t)$ grows only logarithmically, remaining well below the chaotic plateau throughout evolution times of interest, reflecting the slow spread of quantum correlations and the emergence of area-law entanglement. 
The initial state in Aquila is fixed, as the system was originally designed for gate-based quantum computing. Therefore, we focus on studying Hamiltonians whose mid-spectrum energy approximately coincides with the energy of the experimentally accessible initial state. The control parameter governing this energy is $\Delta_{\mathrm{global}}$.
For a more detailed discussion of the role of the initial state and how we design our analysis around it, see Appendix~\ref{ap:magnetisation}.

These distinct dynamical signatures, together with the spectral diagnostics, provide extensive numerical evidence for the qualitatively distinct dynamical regimes at strong and weak disorder in the Rydberg Hamiltonian of Eq.~{\eqref{eqn:hamAquila},
as summarised in Table~\ref{tab:chaos_sig}.

\subsection{Initial state and ground state magnetisation: Staying in the middle of the spectrum}\label{ap:magnetisation}

In this subsection, we explain the physical mechanisms underlying the pronounced phenomenology observed in the Aquila Hamiltonian. The central idea is that the middle of the many-body spectrum must be aligned with the energy of the experimentally accessible initial state, while independently varying the disorder strength. To achieve this, we exploit the role of the global detuning parameter $\Delta_{\mathrm{global}}$, which allows us to shift the spectrum without altering the disorder amplitude.

\begin{figure}
    \centering
    \includegraphics[width=\linewidth]{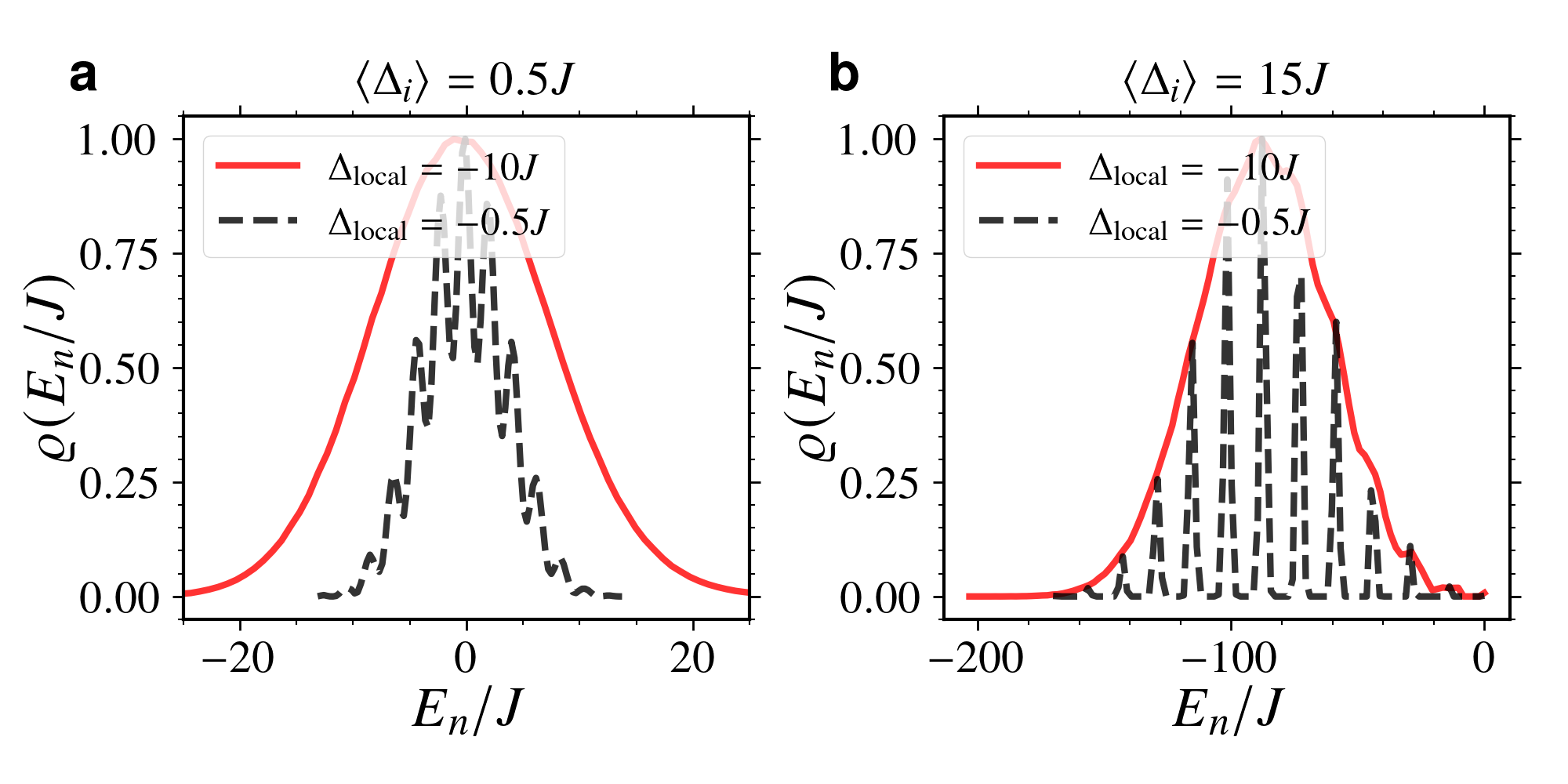}
    \caption{The parameter $\langle \Delta_i \rangle$ induces an overall shift of the many-body spectrum.   
    {\sf{\textbf{a}}} For low values, e.g., $\langle \Delta_i \rangle = 0.5J$, the experimentally accessible initial state—whose energy is zero—lies approximately in the middle of the spectrum.
    {\sf{\textbf{b}}} For large values, e.g., $\langle \Delta_i \rangle = 15J$, the same initial state is located far from the spectral center, effectively outside the main support of the many-body density of states. We use 100 bins.}
    \label{fig:DOS}
\end{figure}

Signatures of quantum chaos and MBL become most pronounced for initial states and eigenstates located near the center of the many-body spectrum, particularly in the scaling behavior of the entanglement entropy and its dynamical growth. In the experimental setting, however, the initial state is constrained to have zero energy [as can be seen by acting with the Hamiltonian of Eq.~(\ref{eqn:hamAquila}) on the state $|g\rangle^{\otimes N}$]. 
To meaningfully probe the transition between chaotic and localised regimes, we therefore consider Hamiltonians whose density of states is centered around zero energy. This is achieved by tuning the average detuning parameter $\langle \Delta_i \rangle = \Delta_{\mathrm{global}} + \Delta_{\mathrm{local}}/2$, thereby aligning the spectral midpoint with the experimentally accessible initial state, see Fig.\,\ref{fig:DOS}.
We further observe that while the central position of the density of states is essentially independent of $\Delta_{\mathrm{local}}$, its structure is strongly affected by disorder. In the weak-disorder regime, the density of states exhibits well-defined peaks associated with distinct magnetisation sectors. As the disorder strength increases, mixing between these sectors smooths out the distribution, resulting in a featureless density of states.

\begin{figure}
    \centering
    \includegraphics[width=\linewidth]{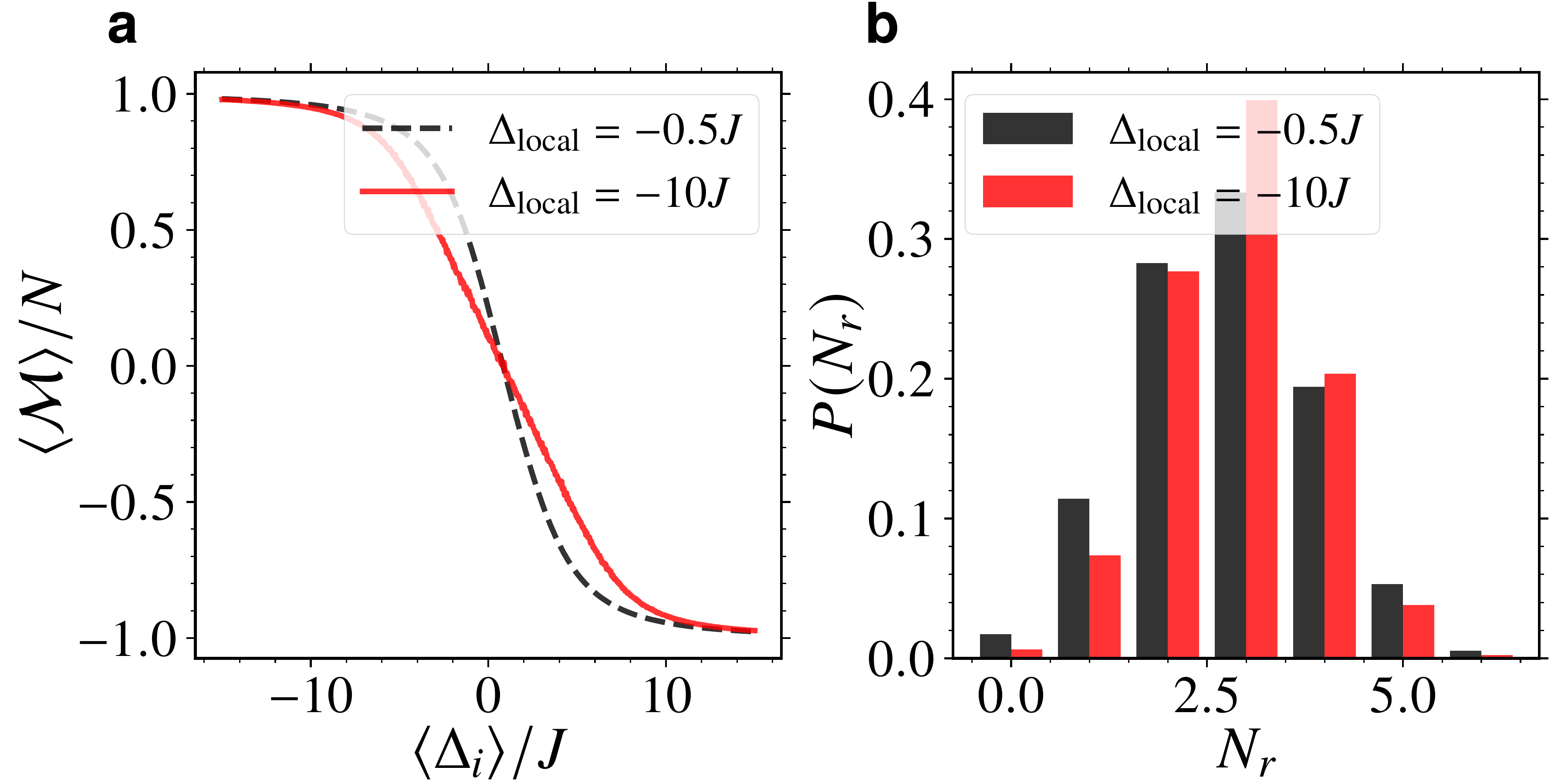}
    \caption{{\sf{\textbf{a}}} Magnetisation of the ground state for an $N=6$  qubit system (normalised by the number of qubits), averaged over 500 disordered realisations , for different amounts of uncertainty in $\Delta_{\mathrm{local}}$.  
    {\sf{\textbf{b}}} Probability of measuring $N_r$ qubits in the Rydberg state for low and high mean $\langle \Delta_{\mathrm{i}} \rangle$, for 6 qubits.}
    \label{fig:2}
\end{figure}

Moreover, the choice of Hamiltonian parameters can also be motivated from the perspective of ground-state magnetisation. The experimentally accessible initial state is fully polarised and therefore has a normalised magnetisation equal to $1$. We thus require a parameter regime in which the ground-state magnetisation differs significantly from that of the initial state. This ensures that the initial state does not lie near the spectral edge, but instead overlaps with eigenstates in the bulk of the spectrum.
In Fig.\,\ref{fig:2}, we present the dependence of the ground-state magnetisation on the average detuning. We show the normalised magnetisation (for both the chaotic and MBL parameters) as a function of the mean detuning $\langle \Delta_i \rangle$. 
We observe that when the average detuning becomes large compared to the interaction energy scale, the ground state progressively approaches a fully polarised configuration.

\begin{figure}
    \centering
    \includegraphics[width=\linewidth]{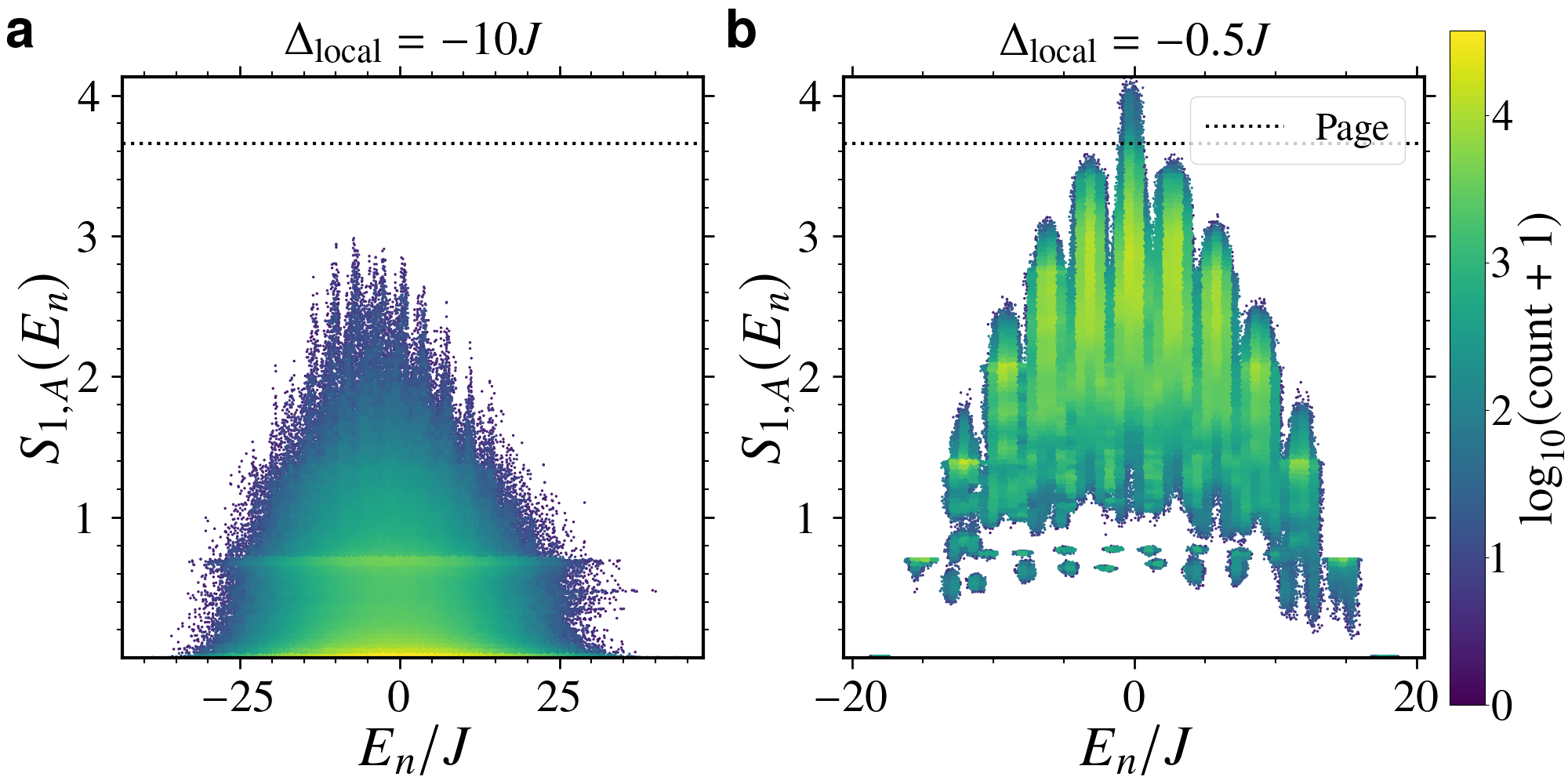}
    \caption{Density plot of the von Neumann entropy of eigenstates for $1000$ realisations of a $N=12$ qubit system with $\langle \Delta_i \rangle = 0.5J$, for an equal bipartition, as a function of the corresponding eigenenergy, $S_{1,A}(E_n)$. In {\sf{\textbf{a}}}, we set $\Delta_{\mathrm{local}}=-10J$ and in {\sf{\textbf{b}}} we set $\Delta_{\mathrm{local}}=-0.5J$. The color scale represents $\log_{10}(\mathrm{count}+1)$ obtained from a $120 \times 120$ two-dimensional histogram, so brighter regions correspond to a higher density of eigenstates at a given energy-entropy pair.
    }
    \label{fig:DEE}
\end{figure}

\subsection{von Neumann Entropy of Eigenstates}
\label{app:S_eigen}

Fig.\,\ref{fig:DEE} shows the distribution of the bipartite von Neumann entropy $S_{1,A}(E_n)$ of energy eigenstates for a $12$-qubit system over $1000$ disorder realisations, plotted as a function of eigenenergy $E_n/J$.
For $\Delta_{\mathrm{local}}=-10J$, see Fig.\,\ref{fig:DEE}{\sf a}, most eigenstates are concentrated at relatively low entropy values, particularly near the spectral edges. Even in the middle of the spectrum, the entropy remains well below the dotted horizontal line indicating the Page value (entropy of typical states). This behavior is characteristic of a MBL regime, where ergodicity is broken and eigenstates exhibit area-law entanglement rather than thermal, volume-law scaling.
In contrast, for $\Delta_{\mathrm{local}}=-0.5J$, see Fig.\,\ref{fig:DEE}{\sf b}, the entropy is maximal near the center of the spectrum and decreases toward the edges. Many mid-spectrum eigenstates approach the Page value, consistent with the ETH and volume-law entanglement. This indicates an ergodic, thermalising regime.

\section{Randomised measurements on Aquila} \label{apx_protocol}
We note that there is an existing protocol in this area: ``A randomised measurement toolbox for an interacting Rydberg-atom quantum simulator" \cite{notarnicola2023}. However, that proposal is not relevant to Aquila for several reasons. The first is that the encoding is based on two Rydberg levels, i.e. $nS$ for $\ket{0}$ and $nP$ for $\ket{1}$ for $n \approx 60$, which leads to a dipole-dipole $\propto 1/r^3$ $\sigma^+_i \sigma^-_i$ interaction. Aquila instead encodes a qubit with a $5S_{1/2}$, $F = 2$ state and a $70S_{1/2}$ state. The second is that their protocol relies on varying a parameter equivalent to Aquila's $h_i$ parameter dynamically, which is not possible for us.

This approach is closely related to the technique of classical shadows \cite{huang2020}. 
In comparison to the classical shadows, randomised measurement is more robust against miscalibration since the exact form of $U_y$ is never used in any calculations, and it is known that with $n_S > n_U$ classical shadows have greater error over randomised measurements \cite{elben2022}. Since it is far cheaper to increase $n_S$ over $n_U$ and to mitigate the effects of miscalibration, we use the postprocessing method of the randomised measurement protocol.

In principle, one could also choose different randomisation processes, e.g., two or more different detuning parameters or a combination of all available controls.
The choice of $\phi = \pi/2$ in this setup is because it can be quenched as a step function between values, whereas all detunings must be changed at a finite maximum rate.

Van Enk and Beenakker \cite{vanenk2012} introduced the idea that randomised measurements could be used to extract estimates for expectation values of polynomial functions of the reduced density matrix. Van Enk and Beenakker consider $\rho^{\mathrm{in}}$, an $M \times M$ reduced density matrix from which we desire to calculate $\mathrm{Tr}((\rho^{in})^n)$. They embed $\rho^{\mathrm{in}}$ inside a new Hilbert space of dimension $N \gg M$ by padding with zeros and apply a random unitary $U \in U(N)$ drawn from the Haar distribution. They argue that the submatrix corresponding to the upper $M\times M$ block will have complex entries whose real and imaginary parts are distributed according to a Gaussian for $N$ large. The moments of the distribution of the measurement outcomes on the $M \times M$ subsystem, with this Gaussian approximation of the entries, obey Isserlis's theorem. They use this observation to derive an expression relating the $n$th order moment of the probability distribution of each computational basis state to $\mathrm{Tr}((\rho^{\mathrm{in}})^n)$. This idea built on earlier work by Beenakker and others where they estimated the purity of two photon states by passing the photons through a random medium and observed the resulting interference pattern \cite{beenaker2009,peeters2010}.

This idea was first tested  experimentally in 2019 by Peter Zoller's group with trapped ions \cite{brydges2019,elben2019}, with two crucial modifications to van Enk and Beenakker's work: instead of sampling from the Haar distribution on a large system to invoke the Isserlis theorem, they use the Weingarten calculus \cite{collins2022} to show that applying unitary matrices sampled from single qubit two-designs, i.e. matching the first two moments of the Haar distribution, is sufficient to calculate the purity. This improvement on the protocol enabled feasible experimental demonstration by drastically reducing the overall size of the system required and removing the need to sample from the full Haar random distribution. 

To get a sense intuitively for how this protocol works, Ref. \cite{brydges2019} gives a good illustration of the case of a single qubit. Imagine a vector on the Bloch sphere. We want to estimate the length of this vector, i.e. the purity, by only projecting onto the polar ($z$) axis. If we can uniformly randomly rotate this vector around the sphere and then project onto the $z$-axis, then the width of the distribution of measurement outcomes yields the length of the original vector. The principle is the same with some $N$ qubit state.

A key ingredient in this demonstration with ions \cite{brydges2019} is single qubit gates. Since the goal of the randomised measurement protocol is to apply random unitary operations on no larger a subsystem than the one of interest---otherwise the purity of the subsystem is changed due to the entanglement generated---having local control seems obviously necessary. However, Aquila does not support single qubit gates. Indeed, as shown in the Hamiltonian in Eq.~\eqref{eqn:hamAquila} the Rabi frequency driving the transition from $\ket{g}\leftrightarrow \ket{r}$ acts globally, i.e. on all qubits simultaneously. What we do have is control over the local field strength $h_i$, so the detuning for each qubit $\Delta_{i} = \Delta_{\mathrm{global}} + h_i \Delta_{\mathrm{local}}$ can be made incommensurate for incommensurate $h_i$ despite having global values for $\Delta_{\mathrm{global}}$ and $\Delta_{\mathrm{local}}$. 

This is sufficient for our purposes. We define two global ``gates" as follows:
\begin{equation}
    U_0 = \exp(-i\tau H_0), \quad U_1 = \exp(-i\tau H_1)
\end{equation}
where
\begin{align}
    H_0 &= \frac{\Omega}{2}\sum_{i=0}^{N-1}\bigg (\hat{\sigma}_i^+ + \hat{\sigma}_i^-\bigg ) - \sum_{i=0}^{N-1} \Delta_{i}\hat{n}_i + \sum_{ i < j} J_{ij}\hat{n}_i\hat{n}_j\\
    H_1 &= \frac{\Omega}{2}\sum_{i=0}^{N-1}\bigg (i\hat{\sigma}_i^+ - i\hat{\sigma}_i^-\bigg ) - \sum_{i=0}^{N-1} \Delta_{i}\hat{n}_i + \sum_{ i < j} J_{ij}\hat{n}_i\hat{n}_j
\end{align}
for fixed values of $\Omega, \Delta_{\mathrm{local}},  \Delta_{\mathrm{global}}, \tau$. Crucially, we can switch essentially instantaneously between $H_0$ and $H_1$ by taking $\phi = 0 \to \frac{\pi}{2}$, which avoids the need to ramp the detuning or Rabi drive. The values $\Delta_{\mathrm{local}},  \Delta_{\mathrm{global}}$ here are not the same as in the ``Preparation" phase of the experiment; we perform optimisation in order to determine how to set them, which we describe momentarily. We sample unitary operations by choosing at random a length $L$ bitstring $y$:
\begin{equation}\label{eqn:product}
    U_y = \prod_{j=1}^L U_{y[j]}.
\end{equation}

This matches the form of the operations considered by Lloyd \cite{lloyd1995, lloyd1996} and Deutsch, Barenco, and Ekert \cite{deutsch1997} where a discrete set of Hamiltonians can be applied in an arbitrary sequence; the resulting time evolution operator is generated by a Hamiltonian in the Lie algebra of that set of Hamiltonians. A full proof for the universality of this choice of gates and a detailed analysis of how the rotations affect different subsystems is the subject of our future work.
The intuition for this choice of gates is that at the single qubit level, $H_0, H_1, [H_0, H_1]$ are linearly independent, meaning they generate $\mathfrak{su}(2)$ on a single qubit in isolation. Extending to multiple qubits but for the moment ignoring any interaction, if each qubit had the same $\Omega, \Delta_i$ then in the same time $t$ they would all undergo the same rotations. While $\Omega$ must remain fixed across the qubits, with local detuning, by selecting random $h_i$ for each qubit, we can ensure that each $\Delta_i$ are  distinct, and hence each qubit experiences a unique rotation in the same time $t$. The question remains how to suppress the multi-qubit interaction term, which must remain present since the atoms cannot be shuttled. Here, the intuition is to exploit the physics that we seek to demonstrate: by localizing the system, even though the qubits interact, the loss of purity is minimised---as shown in Fig.\,\ref{figure1}. 

We choose the parameters $L= 16, \, \tau = 0.06\mu\mathrm{s}, \, \Delta_{\mathrm{local}}= -102.72\mu\mathrm{s}^{-1}, \, \Delta_{\mathrm{global}} = 26.73\mu\mathrm{s}^{-1}$. A large $\Delta_{\mathrm{local}}$ (but random $h_i$) means a large disorder in $\Delta_i$, consistent with our intuition about localisation.
We find these values by optimizing the objective function
\begin{equation}\label{eqn:objective}
    C(\tau, L, \Delta_{\mathrm{global}}, \Delta_{\mathrm{local}}) = w_{\theta,\phi}\overline{\Delta \theta} \overline{\Delta \phi} - w_r \overline{\Delta r}.
\end{equation}
with the weights $ w_{\theta,\phi} = 1.0$ and $w_r = 2.0$. We start with a Haar random state of six qubits and apply 50 independent random rotations of length $L$ defined as in Eq.~\eqref{eqn:product}. For each resulting state, we compute the polar coordinates on the Bloch sphere $(r_{qi}, \theta_{qi}, \phi_{qi})$ for each qubit $q$ after the $i$th rotation, each time starting from the same state. For each value of $q$, we compute $\Delta r_q = \mathrm{max}_i r_{qi} - \mathrm{min}_i r_{qi}$, $\Delta \theta_q = (\mathrm{max}_i \theta_{qi} - \mathrm{min}_i \theta_{qi})/\pi$, $\Delta \phi_q = (\mathrm{max}_i \phi_{qi} - \mathrm{min}_i \phi_{qi})/(2\pi)$. Then, we average over all the qubits to obtain $\overline{\Delta r} = \sum_{q=0}^5 \Delta r_{q}/6$, $\overline{\Delta \theta} = \sum_{q=0}^5 \Delta \theta_{q}/6$, $\overline{\Delta \phi} = \sum_{q=0}^5 \Delta \phi_{q}/6$. 
The first term in Eq.~\eqref{eqn:objective} is a product to ensure that solutions in which both $\overline{\Delta \theta}$ and $ \overline{\Delta \phi}$ are maximal is the optimal solution; i.e., when both the polar and azimuthal angles realise their full range. The second term penalises solutions that cause loss of purity. This simple method captures, at least heuristically, the features desirable for our rotations. 

We use dual annealing from \texttt{scipy.optimise} with 1000 iterations with the initial guess $\tau = 0.05\mu \mathrm{s}, \, \Delta_{\mathrm{global}}=15\mu \mathrm{s}^{-1}, \, \Delta_{\mathrm{local}}=27.5\mu \mathrm{s}^{-1}, \, L=5$ with the constraint $L\tau \leq 1\mu\mathrm{s}$.

We illustrate qualitatively the effect of these choices in Fig.\,\ref{figure2}{\sf a-c}. Fig.\,\ref{figure2}{\sf a} shows the Bloch vector for the third qubit in a chain of 6 atoms. The initial Bloch vector (black arrow) is rotated around the sphere according to either single qubit gates sampled directly from the Haar distribution or our $\phi$-quench process. The final position of the Bloch vector after rotation is shown as rose triangles and blue squares for the Haar and $\phi$-quench processes respectively. A black line in Fig.\,\ref{figure2}{\sf a} starting from the tip of the initial Bloch vector shows the continuous evolution of the Bloch vector under the $\phi$-quench to its final location marked with a black square. 
Below the sphere, we show the distribution of $\langle \sigma_z^i\rangle$ of the final Bloch vectors rotated under local Haar and $\phi$-quench evolutions. While the $\phi$-quench case is more peaked than Haar, it covers the same range of projected values---which is all we need to estimate purity of the original vector. The fact that the distribution has a different shape is explained by Fig.\,\ref{figure2}{\sf b} which shows how the purity of the final Bloch vector changes with the number of choices $\phi$ in the sequence. The average behavior is a solid curve; individual realisations show a fluctuation both above and below the original value.
Fig.\,\ref{figure2}{\textbf{c}} shows how the polar and azimuthal angles densely cover their respective ranges.

\section{Noiseless emulation of the protocol}\label{apx_noiselss_protocol}
\begin{figure}
    \centering
    \includegraphics[width=0.93\linewidth]{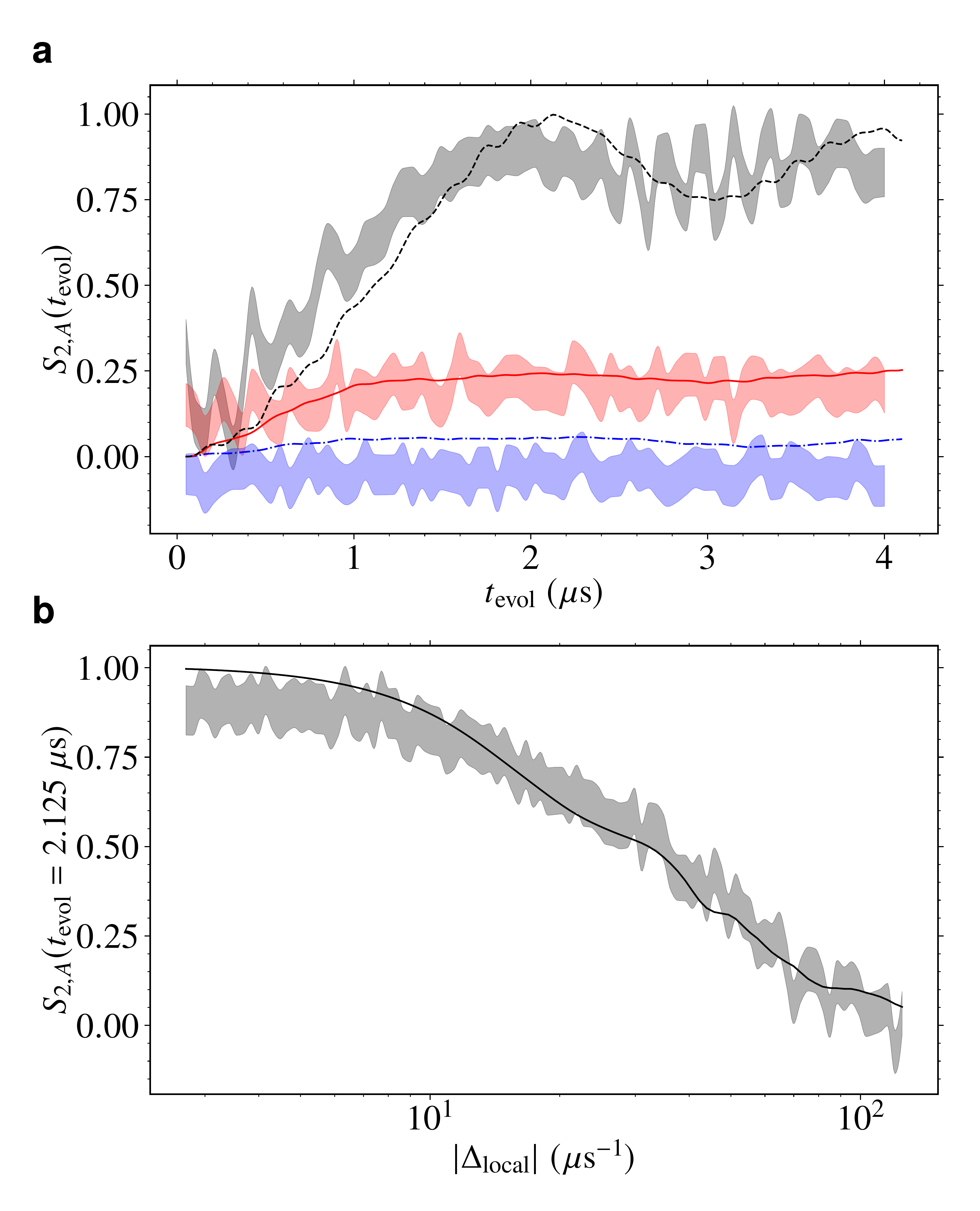}
    \caption{{\sf{\textbf{a}}} Numerically computed second-order R\'enyi entropy as a function of evolution time shown as curves for $\Delta_{\mathrm{local}}=-0.5J$ (black, dashed line), $\Delta_{\mathrm{local}}=-10J $ (red, solid line), and  $\Delta_{\mathrm{local}}=-23.06J $ (blue, dashed-dotted line). The shaded areas show emulated protocols  with $\pm 1\, \mathrm{SEM}$ propagating multinomial shot noise and with the same parameters as  the experiment in Fig.\,\ref{fig:rand}{\sf a}.
    {\sf{\textbf{b}}}  Numerically computed second-order R\'enyi entropy as a function of $\Delta_{\mathrm{local}}$ and emulated protocol  with $\pm 1\, \mathrm{SEM}$ propagating multinomial shot noise and with the same parameters as  the experiment in Fig.\,\ref{fig:rand}{\sf b}.
    }
    \label{fig:bloqade_sim}
\end{figure}

Mirroring Fig.\,\ref{fig:rand}, Fig.\,\ref{fig:bloqade_sim} shows the time and disorder dependence of the second-order R\'enyi entropy.
The curves are obtained by numerically evolving an initial $|gggggg\rangle$ state under the Hamiltonian of Eq.~\eqref{eqn:hamAquila}, including ramps for $\Omega$, $\Delta_{\mathrm{global}}$,  and $\Delta_{\mathrm{local}}$ as illustrated in Fig.\,\ref{fig:rand}{\sf a} but excluding the ``randomise" phase. We directly calculate the R\'enyi entropy values, $S_{2,A}(t_{\mathrm{evol}})$, from reduced density matrices for $A$ the first 3 qubit subsystem. The shaded areas implement the full emulation of the protocol with the same numbers of rotations and disordered realisations of the experiments by including the ``randomise" phase as in the shaded areas of Fig.\,\ref{fig:rand}{\sf a,b}. The shaded areas agree well with the numerical curves, illustrating the success of the protocol in the absence of noise. Note that Eq.~\eqref{eqn:RMTX_renyi} does not prevent the estimated purity from being $>1$, leading to the aphysical negative estimated entropy in the $\Delta_{\mathrm{local}}=-23.06J$ case. However, this is an artifact of using too few $n_U$ and $n_{\mathrm{ens}}$; here we emulate the choices used in the experiment which were made with a finite resource budget.


\section{Experimental Details} \label{apx_experiment}

\begin{table*}[t]
  \centering
  \resizebox{\linewidth}{!}{%
  \begin{tabular}{|c|c|c|c|c|c|}
    \hline
     Step & $t \, (\mu \mathrm{s})$  & $\Omega \, (\mu \mathrm{s}^{-1})$ & $\Delta_{\mathrm{local}}\, (\mu \mathrm{s}^{-1})$ & $\Delta_{\mathrm{global}}\, (\mu \mathrm{s}^{-1})$ & $\phi (\mathrm{rad})$ \\ \hline
    Start $\Delta_{\mathrm{global}}$ & $0$ &$ 0$ & $0$ & $4.065 \; (29.81)\; [65.21]$& $0$\\ \hline
    Prepare $\Delta_{\mathrm{local}}$ & $0.05$ &$ 0$ & $-2.71 \; (-54.2) \; [-125]$ & $4.065 \; (29.81)\; [65.21]$& $0$\\ \hline
    Prepare $\Omega$ & $0.1132$ &$ 15.8$ & $-2.71 \; (-54.2) \; [-125]$ & $4.065 \; (29.81)\; [65.21]$& $0$\\ \hline
    Evolve  & $0.1132 + t_{\mathrm{evol}}$ &$ 15.8$ & $-2.71 \; (-54.2)\;  [-125]$ & $4.065 \; (29.81)\; [65.21]$& $0$\\ \hline
    Ramp & $0.1632 + t_{\mathrm{evol}} $ &$ 15.8$ & $-102.72$ & $26.73$& $0$\\ \hline
    Randomise  & $1.1585 + t_{\mathrm{evol}} $ &$ 15.8$ & $-102.72$ & $26.73$& $0,\pi/2$\\ \hline
    Ramp down $\Omega$ & $1.2217 + t_{\mathrm{evol}} $ &$ 0$ & $-102.72$ & $26.73$& $0$\\ \hline
    Ramp down $\Delta_{\mathrm{local}}$ and Measure & $1.2717 + t_{\mathrm{evol}} $ &$ 0$ & $0$ & $0$& $0$\\ \hline
  \end{tabular}
  }
  \caption{Exact parameter values in $\mu \mathrm{s}^{-1}$ given for the protocol illustrated in Fig.\,\ref{fig:rand}{\sf a}.
  }
  \label{tab:phase_vals}
\end{table*}

\begin{figure}
    \centering
    \includegraphics[width=\linewidth]{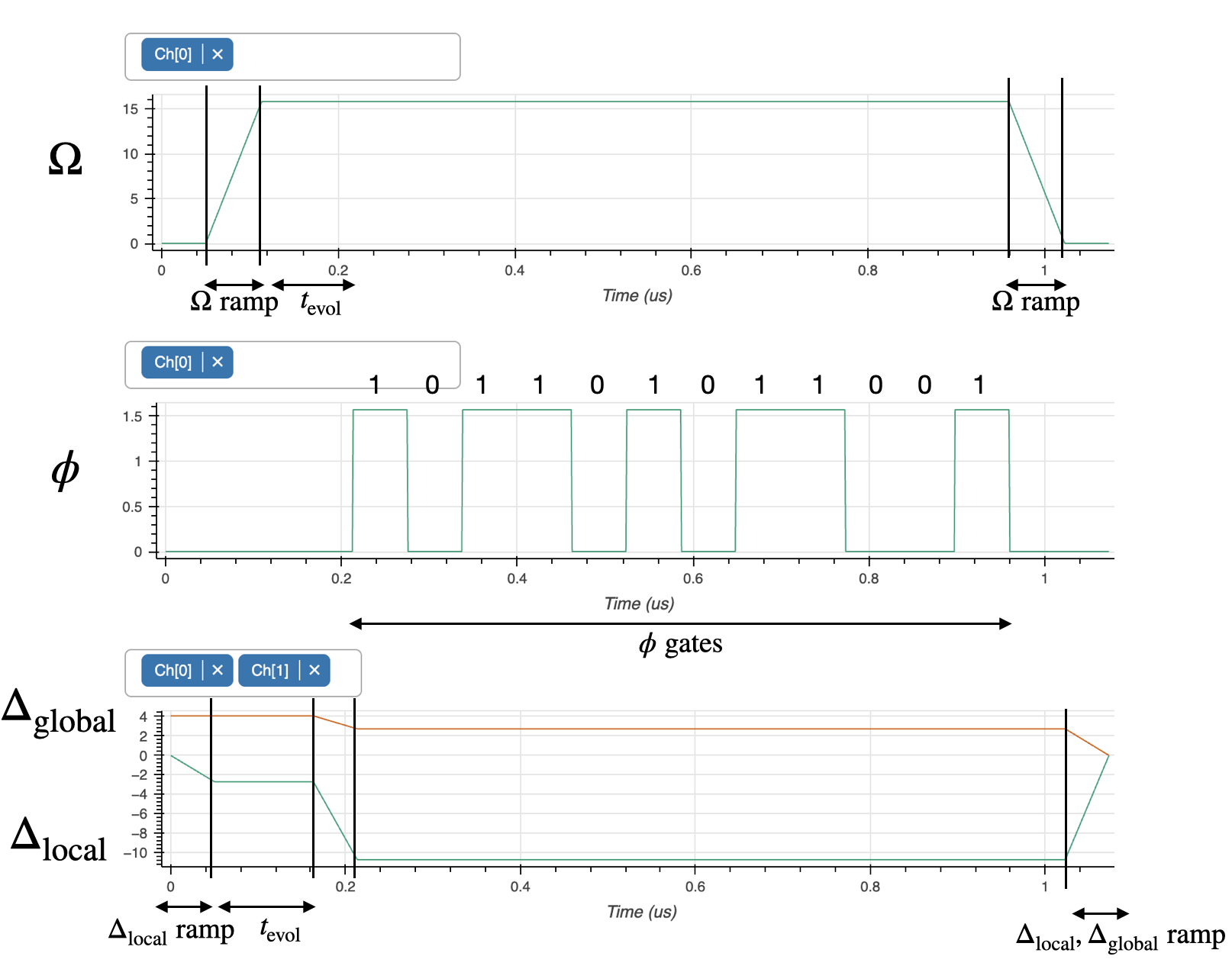}
    \caption{Example task submitted to Aquila showing the time-dependent functions $\Omega, \phi, \Delta_{\mathrm{local}}, \Delta_{\mathrm{global}}$. First $\Delta_{\mathrm{local}}$ is ramped while $\Omega$ is kept at 0 in order to freeze any spurious excitations that can occur when ramping together with $\Delta_{\mathrm{local}}$, then $\Omega$ is ramped up. The system remains constant for a time $t_{\mathrm{evol}}$. $\Delta_{\mathrm{local}}$ and $\Delta_{\mathrm{global}}$ are ramped to their desired values during the randomised phase. $\phi$ flips between $0$ and $\pi/2$ with equal probability for, in the full experiment, 16 times (in this illustration, 12). $\Omega$ is then ramped down to $0$ with $\Delta_{\mathrm{local}}$ still constant. Finally,  $\Delta_{\mathrm{local}}$ and $\Delta_{\mathrm{global}}$ are also ramped down to $0$.}
    \label{fig:ramps_annotated}
\end{figure}

\subsection{Bloqade Analog with Braket}
To perform all the experiments described in this work, we use QuEra's Bloqade Analog Python package (version 0.16.3), which is open source: \url{https://github.com/QuEraComputing/bloqade-analog}. Bloqade Analog has documentation to guides new users and explain all major features of the code. At a high level, Bloqade Analog works by specifying discretised functions for $\Omega(t), \Delta_{\mathrm{global}}(t), \Delta_{\mathrm{local}}(t), \phi(t)$ obeying the minimum timestep of $0.05 \mu \mathrm{s}$ and maximum slew rates for $\Omega$ and $\Delta_{\mathrm{local}},\Delta_{\mathrm{global}}$, in addition to the $(x_i,y_i)$ positions for up to 256 atoms and the local field strengths $h_i$ as lists. Together, these parameters define a \texttt{program} object. 

Programs are executed on Aquila through Braket on AWS. A user has an AWS account which enables them to have access to Braket, which includes a built-in Jupyter notebook. To run a real experiment, the user calls the option \texttt{results = program.braket.aquila().run\_async(shots=n\_shots, use\_experimental=True)}. The boolean \texttt{use\_experimental=True} must be enabled for the user to access local detuning, as it is classified by QuEra as an ``experimental" feature.

Bloqade Analog, when called in experiment mode, includes automatic checks for both the values and rates of change of the values of $\Omega, \Delta_{\mathrm{local}}, \Delta_{\mathrm{global}}$, minimum lattice spacing, and also minimum timesteps of $0.05\mu \mathrm{s}$.

Each time the user executes a command of the form above is called a ``Quantum Task" by AWS. This causes the user to incur a fixed fee of $\$0.3$ plus an additional $\$0.01$ per successful shot. By successful, we mean that due to a high but still finite probability of correctly filling a single lattice position $\approx 99.5\%$, not every shot will correspond to a correctly filled lattice; the user is only charged for those shots that are successful. This task object is recorded on the ``Tasks" page on Braket; it includes timestamps of submission and the status of the task, either ``Queued", ``Running", ``Completed", or ``Failed".
The user must include a line following line executing the program to save it; simply, \texttt{bloqade\_analog.save(results)} which creates a reference file containing the details of the task. When the task has finished running on the hardware, the user can call \texttt{results = bloqade\_analog.load()} to retrieve the reference and then \texttt{results\_data = results.fetch()} to obtain the bitstring measurements.

The user may choose to parallelise their lattice in the experiment. This means that the user will specify a minimum distance between atoms in different ``copies" of their lattice that will be populated to fill the remaining space of the grid. Bloqade Analog will automatically correlate the parallelised lattice sites and return bitstrings as if they are additional shots. We explicitly do not parallelise our lattice because we want to ensure there are no undesired interactions present, since we are working in the non-blockaded regime (see the next section).

\subsection{Calibration checks}

\subsubsection{Rabi oscillations}

We use two primary calibration checks based on Rabi oscillations in order to (1) ensure the Rabi frequency matches what we expect and (2) we understand the readout errors.

We take a chain of 6 atoms spaced $10 \mu \mathrm{m}$ and set $h_i = 1$ except for one value $i'$ for which $h_i' = 0$. We set $\Delta_{\mathrm{local}}=-125 \mu \mathrm{s}^{-1}$. We sweep $i'$ from index 0 to 5 to probe the spatial dependence of the Rabi frequency. We perform the experiment over $0.6 \mu \mathrm{s}$ to capture one period, and run two copies for each detuning pattern: one in which the Rabi oscillation occurs starting at time 0 and starting at $3 \mu \mathrm{s}$ of plateau time to check the effects of decoherence. 

During our experiment, after we measure all the times and rotation sequences for each Hamiltonian,  we sample a single period from 0 to $0.6 \mu \mathrm{s}$ of plateau time of a Rabi oscillation for a single atom to keep a record of the calibration status of Aquila. This is necessary due to the length of time required to run the experiment.

\subsection{Determining the rate of false detection}\label{sec:readout_calibrate}

Using our Rabi experiments as explained in a subsection above, we can fit to the function $f(t) = A\sin(\Omega_{\mathrm{eff}}t + \varphi) + B$, where $A$, $\Omega_{\mathrm{eff}}$, $\varphi$, and $B$ are fit parameters. For a given Rabi oscillation, let the fit parameters for the amplitude and the vertical offset be $A,B$. We use
\begin{align}
    \epsilon_r&= 1-\frac{A+B}{\frac{\Omega^2}{\Omega^2 + \Delta^2}} \label{eqn:e_r}\\
    \epsilon_g&= A-B \label{eqn:e_g}
\end{align}
We use $\Omega, \Delta$ that we have input to Aquila, not fit parameters.

\end{document}